\documentclass[
aps,
physrev,
reprint,
superscriptaddress,
amsmath,
amssymb,
raggedbottom,
nobalancelastpage,
floatfix
]{revtex4-2}

\usepackage[T1]{fontenc}
\usepackage{amsmath,amssymb,bm}
\usepackage{graphicx}
\usepackage{placeins}
\usepackage{booktabs}
\usepackage{array}
\usepackage{xcolor}
\usepackage{siunitx}
\usepackage{microtype}
\usepackage{url}
\usepackage{capt-of}
\usepackage[
    colorlinks=true,
    citecolor=blue,
    linkcolor=blue,
    urlcolor=blue
]{hyperref}
\usepackage{orcidlink}

\newcommand{\kms}{\ensuremath{\mathrm{km\,s^{-1}}}}
\newcommand{\hmpc}{\ensuremath{h\,\mathrm{Mpc}^{-1}}}
\newcommand{\mpch}{\ensuremath{h^{-1}\,\mathrm{Mpc}}}
\newcommand{\kny}{k_{\rm Ny}}

\begin{document}
\title{Initial Conditions and Wave-Phase Resolution for Scalar-Field Dark Matter Simulations}

\author{Hafiz Inam Ullah\,\orcidlink{0009-0005-5796-6054}}
\email{hinamu2500@alumno.ipn.mx}
\affiliation{Escuela Superior de F\'isica y Matem\'aticas, Instituto Polit\'ecnico Nacional, Ciudad de M\'exico, M\'exico}

\author{Muhammad Awais\,\orcidlink{0009-0000-9174-8492}}
\email{mawais2500@alumno.ipn.mx}
\affiliation{Escuela Superior de F\'isica y Matem\'aticas, Instituto Polit\'ecnico Nacional, Ciudad de M\'exico, M\'exico}

\author{Yourong Frank Wang\,\orcidlink{0000-0003-0340-0758}}
\email{yourong.wang@uni-goettingen.de}
\affiliation{Institut f\"ur Astrophysik, Georg-August-Universit\"at G\"ottingen, D-37077 G\"ottingen, Germany}

\author{Tonatiuh Matos\,\orcidlink{0000-0002-0570-7246}}
\email{tonatiuh.matos@cinvestav.mx}
\affiliation{Departamento de F\'isica, Centro de Investigaci\'on y de Estudios Avanzados del IPN, Av. IPN 2508, 07360 Ciudad de M\'exico, M\'exico}

\begin{abstract}
Wave-based cosmological simulations of scalar-field dark matter (SFDM)
require accurate initialization of both the density distribution and
the phase of the complex wavefunction. We develop and validate a matched
$\Lambda$CDM--SFDM initialization framework connecting CLASS-SFDM,
MUSIC, and AxioNyx. For a quadratic SFDM model with
$m_\phi=3.1623\times10^{-22}\,\mathrm{eV}$ at $z_{\rm ini}=120$,
the scalar-component half-power scale is
$k_{50}^{\phi}\simeq10.54\,h\,\mathrm{Mpc}^{-1}$.
An independent AxiECAMB comparison yields a normalized RMS deviation
of $0.328\%$ in the synchronous-gauge SFDM velocity divergence over
$0.20944\leq k\leq13.4\,h\,\mathrm{Mpc}^{-1}$.
A $256^3$ MUSIC realization reproduces the CLASS density power spectrum
over $8\leq k\leq13.404\,h\,\mathrm{Mpc}^{-1}$ with a median ratio
of $1.00131$ and an RMS fractional deviation of $0.138\%$.
Despite this agreement, full-volume phase diagnostics reveal that
the fraction of neighbouring phase increments exceeding the Nyquist
limit decreases from $46.7\%$ at $256^3$ to $13.685\%$ at $512^3$
and $0.124060\%$ at $1024^3$, demonstrating nearly resolved phase
sampling at the highest resolution at initialization.
A separately generated realization yields corresponding violation
fractions of $13.7299\%$ and $0.124983\%$ at $512^3$ and $1024^3$,
respectively, with relative differences below $1\%$ between
the two realizations.
In a common-parent $512^3$--$1024^3$ comparison, the central-plane
density fields remain strongly correlated ($r_\delta=0.9989$).
The median absolute circular phase difference is $0.264$ rad
without masking and $0.246$ rad for the $86.6\%$ of blocks
satisfying $R\geq0.5$, with corresponding 90th percentiles of
$0.826$ and $0.753$ rad.
These results establish that accurate recovery of linear density
perturbations alone is insufficient to guarantee adequate wave-phase
resolution and provide a quantitative framework for validating
SFDM initial conditions prior to nonlinear evolution.
\end{abstract}

\maketitle

\section{Introduction}
\label{sec:introduction}

The standard $\Lambda$CDM cosmological model provides a successful
framework for describing the expansion history of the Universe and
the formation of cosmic structure. Within this framework, cold dark
matter (CDM) drives the gravitational growth of structure, while the
cosmological constant accounts for the accelerated expansion at late
times. Together with baryons, radiation, and neutrinos, these
components determine the evolution of cosmological perturbations
from the early Universe to the present epoch. The increasing precision
of cosmological observations, however, places correspondingly stringent
requirements on theoretical predictions and numerical simulations.
In particular, the initial conditions supplied to nonlinear
calculations must reproduce the intended cosmological perturbations
while controlling errors introduced by numerical approximations.

Linear Boltzmann solvers, including CLASS and CAMB, provide the
species-dependent density and velocity transfer functions used
to generate cosmological initial conditions
\citep{MaBertschinger1995,SeljakZaldarriaga1996,
LewisChallinorLasenby2000,Lesgourgues2011,Blas2011}.
The subsequent evolution is typically followed using particle-based,
hydrodynamical, or wave-based numerical methods
\citep{AnguloHahn2022}. Ensuring consistency between these stages
is essential, particularly when different matter components exhibit
distinct perturbation histories. Approximations involving
scale-independent growth, total-matter transfer functions, or
inconsistent density and velocity conventions can introduce
systematic differences in the initialized fields
\citep{HahnAbel2011,ValkenburgVillaescusa2017,Crocce2006,Michaux2021}.
Consequently, the accuracy of the initialization procedure must
be assessed independently of the subsequent nonlinear evolution.

Scalar-field dark matter (SFDM) provides a well-motivated
alternative description of the cosmological dark matter component
\citep{Matos:1998vk,MatosUrena2001,Matos2009,Matos:2023usa,Awais2026ULDM}.
In the ultralight regime, commonly associated with fuzzy or axion-like
dark matter, the scalar field behaves approximately as CDM on
sufficiently large scales, while wave effects suppress the growth
of perturbations below a characteristic mass-dependent scale
\citep{PressRydenSpergel1990,Peebles2000,Hu2000,Hlozek2015,
Hui2017,Niemeyer2020,Ferreira2021}.
Accurate predictions of this suppression require careful treatment
of scalar-field perturbations, particularly during the transition
to rapid background oscillations. Several numerical formulations
and effective-fluid approaches have been developed to describe
this regime
\citep{UrenaGonzalez2016,UrenaLopez2024CLASS,
PassagliaHu2022,LiuHuGrin2025}.

Beyond linear perturbation theory, SFDM exhibits wave interference
and spatial structure that cannot be represented directly by a
conventional collisionless-particle description. Cosmological
simulations based on the Schr\"odinger--Poisson equations and
related wave-aware approaches have therefore become important
tools for investigating its nonlinear behavior
\citep{WidrowKaiser1993,Schive2014Nature,Schive2014PRL,
Schwabe2016,VeltmaatNiemeyer2016,Mocz2017,Veltmaat2018,
NoriBaldi2018,Mocz2019,MaySpringel2021,Lague2024}.
These developments make the construction of reliable wave initial
conditions a necessary step in connecting linear SFDM predictions
to nonlinear structure formation.

Unlike particle-based initial conditions, a wave representation
requires the initialization of a complex field whose amplitude
and phase encode the density and velocity distributions,
respectively. The peculiar velocity is related to the spatial
gradient of the wavefunction phase. Consequently, a numerical
grid may reproduce the linear density power spectrum accurately
while failing to resolve the phase increments required by the
underlying velocity field. This limitation is particularly
relevant for ultralight scalar fields, where the representable
velocity range depends on the particle mass, initialization
redshift, box size, and spatial resolution. The phase-Nyquist
condition therefore introduces a numerical requirement distinct
from the conventional Fourier-space assessment of density
accuracy. Moreover, adequate sampling of the intended phase
increments does not, by itself, establish convergence of the
reconstructed wave current or subsequent nonlinear evolution.

These issues are relevant to applications involving small-scale
cosmological structure. The Lyman-$\alpha$ forest, which traces
absorption by neutral hydrogen in the intergalactic medium (IGM),
provides constraints on models with suppressed small-scale
clustering, including ultralight dark matter
\citep{GunnPeterson1965,Rauch1998,HuiGnedin1997,
GnedinHui1998,Irsic2017,Armengaud2017,Kobayashi2017,
Nori2019,RogersPeiris2021}.
However, interpreting the transmitted-flux statistics requires
modeling the baryonic density and velocity fields, the thermal
history of the IGM, pressure smoothing, and redshift-space
effects
\citep{Viel2008,Viel2013,Becker2011,Walther2019,
Arinyo2015,PalanqueDelabrouille2015,Chabanier2019}.
Recent work on mixed wave and particle descriptions further
motivates careful treatment of the underlying density and
velocity fields \citep{Wang2026}. Reliable SFDM initial conditions
are therefore a prerequisite for such calculations, although
their validation alone does not constitute a prediction of
Lyman-$\alpha$ observables.

In this work, we develop and validate a matched
$\Lambda$CDM--SFDM initialization framework connecting
CLASS-SFDM, MUSIC, and AxioNyx. We examine the consistency
of the component-specific linear transfer functions, including
an independent AxiECAMB comparison, and assess the numerical
conversion of density and velocity perturbations into the
fields required by MUSIC. We then test the recovery of the
linear density power spectrum and investigate the spatial
sampling of the resulting complex wavefunction. In particular,
we combine phase-Nyquist diagnostics with common-parent
resolution comparisons and full-volume measurements of
neighboring phase increments to quantify the distinction
between density recovery and wave-phase resolution.

The objective is to establish a reproducible validation
framework for cosmological SFDM initial conditions and to
identify the numerical requirements that must be satisfied
before nonlinear wave evolution is undertaken. The analysis
is restricted to the initialization stage; convergence of
the subsequent nonlinear dynamics and predictions for
Lyman-$\alpha$ forest observables are beyond the scope
of the present study. Numerical inputs, source modifications,
and validation products are documented to support reproducibility
\citep{Peng2011,Sandve2013,Wilson2014,Wilkinson2016}.
In particular, the independent velocity-transfer comparison
and full-volume phase diagnostics provide complementary
tests of the initialization pipeline. These results distinguish
the accuracy of the prescribed linear perturbations from the
spatial resolution required to represent the SFDM wavefunction.

\section{Matched initial-condition strategy}
\label{sec:hierarchy}

We organize the calculation into four logically distinct layers:

\begin{enumerate}
\item \textit{Linear theory:} does the Boltzmann solver produce a stable,
well-defined scalar-density and velocity target?
\item \textit{Initial-condition interface:} are density and velocity transfer
functions passed to MUSIC with the correct normalization, sign, units, and
gauge convention?
\item \textit{Wave construction:} does the complex wavefunction reproduce the
target density while resolving the phase gradients implied by the velocity?
\item \textit{Nonlinear evolution:} are the subsequent wave dynamics
converged in space and time?
\end{enumerate}

We validate the linear target and its realization, then probe phase sampling;
nonlinear evolution is outside the scope of this study. Agreement in realized
density power is therefore not treated as evidence of resolved wave phase.

The numerical pipeline employed in this work is
CLASS-SFDM $\rightarrow$ signed transfer tables
$\rightarrow$ MUSIC v1.53 $\rightarrow$ AxioNyx.
CLASS computes the linear, species-dependent density
and velocity transfer functions within the relativistic
Boltzmann framework
\citep{MaBertschinger1995,SeljakZaldarriaga1996,
LewisChallinorLasenby2000,Lesgourgues2011,Blas2011,
LesgourguesTram2011}.
The resulting transfer functions are supplied to MUSIC,
the multiscale initial-condition generator developed by
Hahn and Abel \citep{HahnAbel2011}, to construct matched
cosmological density and velocity realizations.
These initial conditions are subsequently passed to
AxioNyx, the wave-based simulation framework introduced
by Schwabe et al.\ \citep{Schwabe2020}, which extends
the Nyx cosmological code \citep{Almgren2013}
to evolve ultralight scalar dark matter using the
Schr\"odinger--Poisson equations alongside a particle
component.
Recent applications of AxioNyx to mixed dark matter
models provide further motivation for the present
initial-condition validation
\citep{Lague2024,Wang2026}.

\section{Physical model and phase-resolution criterion}
\label{sec:model}

We consider a minimally coupled real scalar field
with a quadratic potential,
\citep{MatosGuzmanUrena2000,MatosUrena2001,Matos2009},
and related wave-dark-matter treatments
\citep{WidrowKaiser1993,Hui2017,Mocz2017},
\begin{equation}
V(\phi)=\frac{1}{2}m_\phi^2\phi^2,
\end{equation}
with fiducial mass
\begin{equation}
m_\phi=10^{-21.5}\,\mathrm{eV}
      =3.16227766\times10^{-22}\,\mathrm{eV}.
\end{equation}
The modified CLASS implementation follows the angular-variable strategy of
\citet{UrenaGonzalez2016} for the rapidly oscillating background and
linear perturbations.

In the nonrelativistic wave description we write the comoving wavefunction as
\begin{equation}
\Psi(\bm x,a)
=
\sqrt{\rho_{\phi,c}(\bm x,a)}\,e^{i\theta(\bm x,a)},
\end{equation}
with $\rho_{\phi,c}=a^3\rho_{\phi,\rm phys}$. The wave phase is related to the peculiar velocity through
equation~(\ref{eq:vphase}) \citep{WidrowKaiser1993,Hui2017,Mocz2017}.
\begin{equation}
\bm v_\phi
=
\frac{\hbar}{a m_\phi}\nabla_{\bm x}\theta.
\label{eq:vphase}
\end{equation}
For neighboring samples separated by $\Delta x$ along direction $\alpha$,
\begin{equation}
\Delta\theta_\alpha^{(v)}
\simeq
\frac{a m_\phi\Delta x}{\hbar}\,
v_{\alpha,\rm link}.
\label{eq:dtheta}
\end{equation}
Because the phase is sampled modulo $2\pi$, an unwrapped nearest-neighbor
increment cannot be represented uniquely once
$|\Delta\theta|>\pi$.  The corresponding phase-Nyquist velocity is
\begin{align}
v_{\rm Ny}^{\rm phase}
&=
\frac{\pi\hbar}{a m_\phi\Delta x}
=
\frac{\pi\hbar N}{a m_\phi L},
\nonumber\\
&=
13.2547\,\kms
\left(\frac{N}{256}\right)
\left(\frac{30\,\mpch}{L}\right)
\nonumber\\
&\quad\times
\left(\frac{3.16227766\times10^{-22}\,\mathrm{eV}}{m_\phi}\right)
\left(\frac{1+z}{121}\right).
\label{eq:phase_nyquist}
\end{align}
We also use $|\Delta\theta|=\pi/2$ as a deliberately stricter diagnostic,
corresponding locally to four samples per $2\pi$ phase cycle.

\begin{table}
\centering
\caption{\label{tab:phase_ny}
Phase-Nyquist velocity for the fiducial
$m_\phi=3.16227766\times10^{-22}\,\mathrm{eV}$,
$L=30\,\mpch$, $z_{\rm ini}=120$ setup.}
\begin{tabular}{rc}
\toprule
$N$ & $v_{\rm Ny}^{\rm phase}$ [\kms]\\
\midrule
64   & 3.3137\\
128  & 6.6274\\
256  & 13.2547\\
512  & 26.5094\\
1024 & 53.0188\\
\bottomrule
\end{tabular}
\end{table}

Equation~(\ref{eq:phase_nyquist}) is useful before any wave evolution is
attempted: density modes can be well represented in Fourier space while the
velocity-implied phase gradient is already aliased.

\section{Linear inputs and reproducibility}
\label{sec:linear}

\subsection{Fiducial cosmology and technical regulator}

The intended physical model contains baryons plus a pure SFDM dark sector.
The fiducial CLASS-SFDM calculation uses
$H_0=67.4\,\mathrm{km\,s^{-1}\,Mpc^{-1}}$,
$\omega_b=0.0224$, $n_s=0.96605$,
$\ln(10^{10}A_s)=3.0448$, $N_{\rm ur}=3.046$,
$N_{\rm ncdm}=0$, and $z_{\rm ini}=120$
\citep{Lesgourgues2011,Blas2011,UrenaGonzalez2016,UrenaLopez2024CLASS}.
The scalar run contains
$\Omega_{\rm sfdm}=0.264$ and a technical synchronous-gauge
$\Omega_{\rm cdm}=10^{-4}$ regulator; the matched $\Lambda$CDM reference
uses $\Omega_{\rm cdm}=0.2641$.

The regulator is not interpreted as physical CDM.  Reducing it from
$10^{-4}$ to $10^{-5}$ and $10^{-6}$ while holding the total dark-sector
abundance fixed changes the total-matter power over the adopted cutoff/IC
interval by only $0.0170\%$ and $0.00166\%$ in the median between successive
steps.  The direct $10^{-4}\rightarrow10^{-6}$ comparison gives a median
difference of $0.0187\%$ and a maximum of $0.1335\%$ over that interval.

The exact scalar settings used in the fiducial calculation are
\begin{verbatim}
attractor_ic_sfdm_1 = yes
sfdm_parameters_1   = -21.5, 0.0, 1., 3., 1.e-30
sfdm_tuning_index_1 = 2
\end{verbatim}
with CLASS 
\citep{UrenaGonzalez2016,UrenaLopez2024CLASS}.

The modified CLASS build exposes the SFDM controls listed above and does not
provide a separate user-facing input parameter explicitly labeled as a
cutoff, transition, or SFDM tolerance setting. The surviving modified
CLASS \citep{UrenaGonzalez2016,UrenaLopez2024CLASS} source nevertheless defines an internal angular transition,
with reference threshold $\theta_*=100$ and sharpness
$\theta_{\rm tol}=1$, as tested below. These source-level settings must
not be conflated with a demonstrated reconstruction of every internal choice
in the original CLASS \citep{UrenaGonzalez2016,UrenaLopez2024CLASS} production runs. We archive the available
configuration and source-provenance information and distinguish the original
transfer tables from the later controlled threshold experiment.

\subsection{Matter-transfer definitions and regulator convergence}
\label{sec:cutoff_regulator_convergence}

The cutoff scale depends on the quantity used to characterize
power suppression. We therefore distinguish the suppression of
the scalar-field component from that of the explicitly
reconstructed physical matter density. For the latter, we
construct the matter transfer function using the baryon and
scalar-field contributions, excluding the technical CDM
regulator from the density weights. The regulator remains
present in the underlying CLASS solution.

We define the normalized power ratio as
\begin{equation}
\mathcal{R}_{P}(k)
=
\frac{P_{\mathrm{SFDM}}(k)/P_{\Lambda\mathrm{CDM}}(k)}
{\operatorname{median}_{\,0.01\leq k\leq 0.1}
\left[
P_{\mathrm{SFDM}}(k)/P_{\Lambda\mathrm{CDM}}(k)
\right]},
\label{eq:normalized_power_ratio}
\end{equation}
where the normalization interval is expressed in
$h\,\mathrm{Mpc}^{-1}$. For each threshold
$x\in\{0.90,0.75,0.50,0.25\}$, we define $k_x$ as the first
downward crossing satisfying
\begin{equation}
\mathcal{R}_{P}(k_x)=x.
\label{eq:cutoff_definition}
\end{equation}
The component-level ratio uses the SFDM scalar-density power in the
numerator and the matched $\Lambda$CDM CDM-density power in the denominator;
the physical-matter ratio instead compares the consistently constructed
physical matter spectra of the two cosmologies. These are separate cutoff
definitions and are not interchangeable. The respective first downward
crossings are determined by interpolation in $\ln k$.

We also test the sensitivity of the inferred cutoff positions
to the technical CDM regulator. We consider
$\Omega_{\mathrm{cdm}}=10^{-4},10^{-5},10^{-6}$, compensating
the change in the regulator density by adjusting
$\Omega_{\phi}$ so that
$\Omega_{\mathrm{cdm}}+\Omega_{\phi}=0.2641$ remains fixed.
All three configurations are analyzed using the same
normalization interval and cutoff-crossing prescription.

\subsection{Stability of the inferred cutoff scales}
\label{sec:cutoff_stability}

The cutoff positions depend weakly on the adopted matter-transfer
definition. For the explicitly reconstructed physical-matter
spectrum, we obtain
$k_{90}=6.744705\,h\,\mathrm{Mpc}^{-1}$,
$k_{75}=8.583279\,h\,\mathrm{Mpc}^{-1}$,
$k_{50}=10.538695\,h\,\mathrm{Mpc}^{-1}$, and
$k_{25}=12.295859\,h\,\mathrm{Mpc}^{-1}$.
The corresponding direct CLASS power-spectrum estimates agree
within approximately $0.0033\%$ in the cutoff positions.

The regulator-convergence test provides an additional check.
When $\Omega_{\mathrm{cdm}}$ is reduced from $10^{-4}$ to
$10^{-6}$ while the total dark-sector background density is
held fixed, the physical-matter cutoff positions shift by less
than $0.011\%$ for all four thresholds. The component-level
cutoff positions exhibit similarly small shifts. These results
demonstrate the stability of the measured cutoff positions
under the tested regulator variations. They do not, on their
own, establish convergence of the full transfer functions at
all wavenumbers.

\subsection{Internal oscillation-transition threshold sensitivity}
\label{sec:theta_sensitivity}

In the surviving modified CLASS implementation, the rapidly
oscillating trigonometric contributions are multiplied by the
internal window
\begin{equation}
W(\theta)=\frac12\left[1-\tanh\!\left(
\theta_{\rm tol}(\theta^2-\theta_*^2)\right)\right],
\end{equation}
with $\theta_{\rm tol}=1$ and reference $\theta_*=100$.
We performed four controlled calculations with
$\theta_*\in\{50,100,200,400\}$, keeping the cosmological
parameters, transfer sampling, and other numerical settings
fixed. These tests were performed using the surviving
CLASS source.

The scalar-component half-power scale was measured from the
first downward crossing of the low-$k$-normalized squared
density-transfer ratio, using the same
$0.01\leq k\leq0.1\,h\,\mathrm{Mpc}^{-1}$ normalization
interval for all four calculations. The resulting values are
listed in Table~\ref{tab:theta_sensitivity}.

\begin{table}
\centering
\caption{Sensitivity of the scalar-component half-power scale
to the internal oscillation-transition threshold in the
surviving CLASS~3.3.4 implementation. Shifts are relative to
$\theta_*=100$, and $k_{50}^{\phi}$ is in
$h\,\mathrm{Mpc}^{-1}$.}
\label{tab:theta_sensitivity}
\begin{tabular}{rrr}
\toprule
$\theta_*$ & $k_{50}^{\phi}$ & Shift [\%]\\
\midrule
50  & 10.555470 & +0.189949\\
100 & 10.535458 & 0.000000\\
200 & 10.521244 & -0.134916\\
400 & 10.513508 & -0.208344\\
\bottomrule
\end{tabular}
\end{table}

The successive relative shifts for $50\to100$, $100\to200$,
and $200\to400$ are $-0.189589\%$, $-0.134916\%$, and
$-0.073527\%$, respectively. Their decreasing magnitudes
indicate progressive stabilization of the
\emph{half-power location} over the tested range, rather than
exact independence of the transition prescription. In
particular, the maximum pointwise relative difference in the
normalized suppression ratio for $\theta_*=400$ versus
$\theta_*=100$ is $0.888410\%$ over
$8\leq k\leq13.4\,h\,\mathrm{Mpc}^{-1}$. Thus, although the
half-power location changes by only a small fraction of a
percent, the full transfer curve is not demonstrated to
converge to the same tolerance.

\subsection{Background evolution at the common transition threshold}
\label{sec:theta_background_mapping}

As an additional check, we examine the background evolution
of the angular variable $\theta_{\rm SFDM}$ in the four
controlled CLASS calculations. The background output
provides $\theta_{\rm SFDM}$ as a function of redshift.
For each calculation, we determine the first crossing of
the common diagnostic level
$|\theta_{\rm SFDM}|=100$, using interpolation in $\ln a$
between the neighboring background samples. This comparison
tests whether the background evolution at a common angular
value is sensitive to the different internal transition
thresholds.

The resulting crossing redshifts are
$z_{\theta=100}=4.72781\times10^5$,
$4.72720\times10^5$,
$4.72721\times10^5$, and
$4.72722\times10^5$ for
$\theta_*=50$, 100, 200, and 400, respectively.
Relative to the $\theta_*=400$ calculation, the maximum
difference is $0.012481\%$, occurring for $\theta_*=50$.
For $\theta_*=100$ and 200, the corresponding differences
are $-0.000423\%$ and $-0.000212\%$, respectively.

\begin{table}
\centering
\caption{Redshift at which the background angular variable
first reaches the common diagnostic level
$|\theta_{\rm SFDM}|=100$ in the four controlled
CLASS~3.3.4 calculations. Relative differences are measured
with respect to the $\theta_*=400$ calculation.}
\label{tab:theta_background_mapping}
\begin{tabular}{rrr}
\toprule
$\theta_*$ & $z_{\theta=100}$ & Difference [\%] \\
\midrule
50  & $4.72781\times10^5$ & +0.012481 \\
100 & $4.72720\times10^5$ & -0.000423 \\
200 & $4.72721\times10^5$ & -0.000212 \\
400 & $4.72722\times10^5$ & 0.000000 \\
\bottomrule
\end{tabular}
\end{table}

The small differences in the crossing redshift indicate that
the background evolution at this common diagnostic level is
stable across the tested transition prescriptions. Combined
with the half-power comparison above, these results provide
complementary evidence for the numerical stability of the
background crossing and the inferred cutoff location.
However, this conclusion is restricted to the tested
background crossing and half-power measurements; it does not
establish convergence of the complete background and
perturbation solutions at all redshifts and wavenumbers.
\subsection{Signed transfer functions}

\begin{figure}
\centering
\includegraphics[width=0.98\columnwidth]{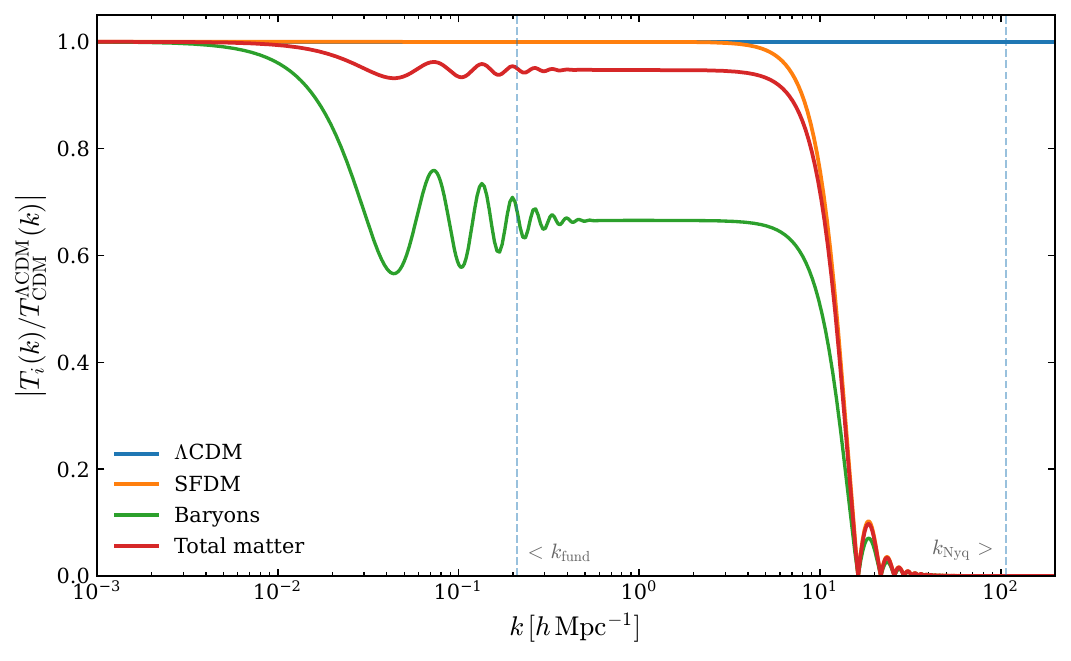}
\caption{Linear transfer amplitudes at $z_{\rm ini}=120$, normalized to the CDM transfer of the matched $\Lambda$CDM reference. The SFDM component follows the $\Lambda$CDM behavior on large scales but develops a pronounced small-scale suppression and oscillatory structure around and beyond the characteristic cutoff scale. The plotted total-matter curve is the density-weighted CLASS-SFDM matter combination. The technical CDM regulator is included only in the CLASS bookkeeping and is not interpreted as a physical CDM component}.
\label{fig:transfer_components}
\end{figure}

The dense transfer tables are evaluated directly at $z=120$ with
$P_{k,\max}=200\,h\,\mathrm{Mpc}^{-1}$.  Fig.~\ref{fig:transfer_components}
shows the component hierarchy normalized to the CDM transfer of the matched
$\Lambda$CDM calculation.  We retain transfer-function signs in the
CLASS-to-MUSIC interpolation because the scalar and baryon transfers become
oscillatory at high $k$.  Clipping negative values would change Fourier
phases and is therefore not an innocuous numerical choice.
To accommodate signed SFDM transfer functions, we use a modified
MUSIC \texttt{camb\_file} reader based on Git revision
\texttt{6248d13}. The modification in
\texttt{src/plugins/transfer\_camb.cc} preserves the signs of
density and velocity transfer functions by avoiding logarithmic
transformations of non-positive-definite transfer columns.
The corresponding signed interpolation and extrapolation
preserve the oscillatory structure across transfer-function
zero crossings. Independent regeneration of the archived
$256^3$ SFDM and baryonic-particle MUSIC initial conditions
reproduces all ten native output fields exactly in each case,
confirming the numerical reproducibility of the implemented
transfer interface.

\begin{figure*}
\centering
\includegraphics[width=0.96\textwidth]{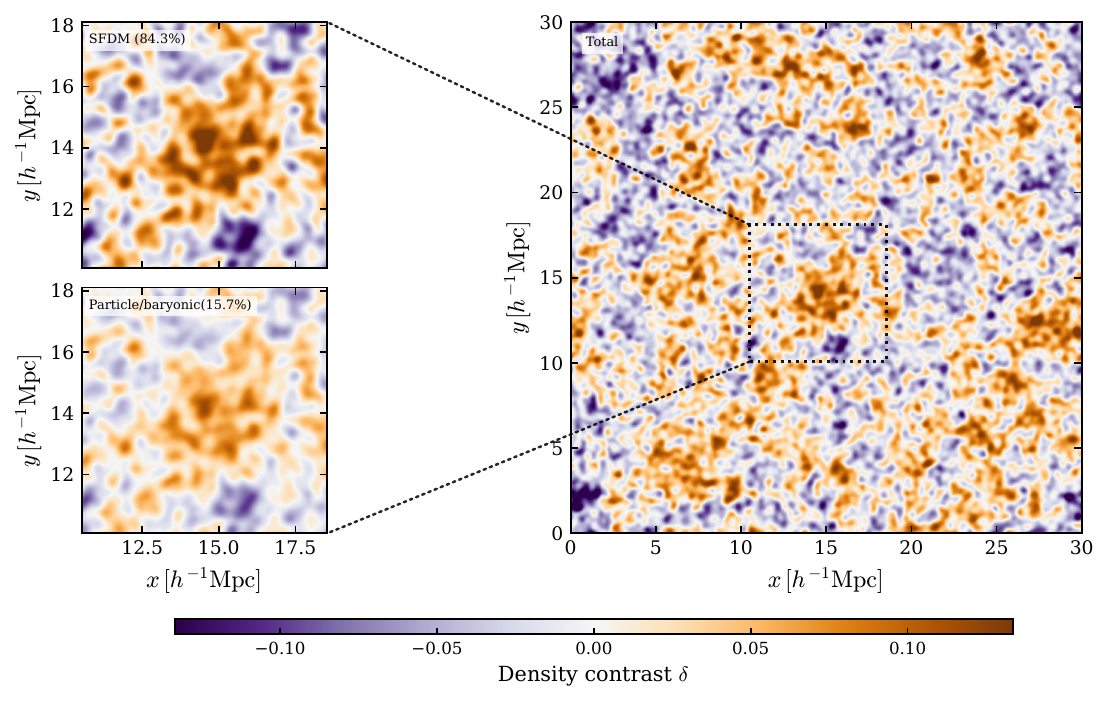}
\caption{Central-plane density contrast of the $1024^3$ initial conditions at
$z_{\rm ini}=120$.  The upper-left panel shows the SFDM component, which
contributes $84.3\%$ of the physical matter density, and the lower-left
panel shows the particle/baryonic component, which contributes $15.7$ per
cent.  The right panel shows the corresponding total density contrast across
the full $30\,h^{-1}\,\mathrm{Mpc}$ box.  Dashed lines identify the subregion
enlarged in the component panels.  The technical CLASS CDM regulator is
excluded from these physical fractions.  The figure illustrates the component decomposition of the initialized field;
quantitative consistency and resolution tests are presented in the following
sections.}
\label{fig:component_density}
\end{figure*}

The transfer-space analysis is complemented by a real-space inspection of the
initialized components.  Excluding the technical CDM regulator from the
physical matter budget, the fiducial model has
$f_{\phi}=\Omega_{\phi}/(\Omega_{\phi}+\Omega_b)=0.8426$ and
$f_b=0.1574$.  Fig.~\ref{fig:component_density} shows a representative
central plane of the $1024^3$ initial conditions.  The SFDM and
particle/baryonic fields trace the same large-scale realization but differ in
their small-scale morphology, and their density-weighted combination gives
the total field shown over the full box.  This figure provides a real-space view of the component decomposition; the
quantitative Fourier- and phase-space tests follow below.

The CLASS \texttt{d\_tot} output includes radiation at $z=120$.
Whenever a true matter-only quantity is needed we therefore construct
\begin{equation}
T_{m,\mathrm{phys}}^{\Lambda\mathrm{CDM}}(k)
=
\frac{
\Omega_b T_b^{\Lambda\mathrm{CDM}}(k)
+
\Omega_{\mathrm{cdm}}^{\Lambda\mathrm{CDM}}
T_{\mathrm{cdm}}^{\Lambda\mathrm{CDM}}(k)
}{
\Omega_b+\Omega_{\mathrm{cdm}}^{\Lambda\mathrm{CDM}}
}.
\label{eq:physical_lcdm_transfer}
\end{equation}
\subsection{Sensitivity to initial-condition transfer prescriptions}
\label{subsec:transfer_prescriptions}

We examine the sensitivity of the SFDM and baryonic
transfer functions to alternative prescriptions used
in constructing cosmological initial conditions.
Specifically, we compare the component-specific CLASS
reference with a longitudinal-gauge velocity
reconstruction, scale-independent rescaling of
transfer functions from $z=0$, and replacement of
component-specific transfers by total-matter transfers.

These comparisons probe distinct assumptions about
gauge conventions, redshift evolution, and component
dependence. Differences near zeros of the reference
transfer functions require particular care because
small absolute deviations can produce large
fractional ratios.

The detailed prescriptions, numerical comparisons,
and associated figure are presented in
Appendix~\ref{app:transfer_prescriptions}.
These diagnostics complement the component-specific
CLASS-to-MUSIC validation but do not directly measure
errors introduced by MUSIC or AxioNyx.

\subsection{Density and velocity conventions}

For species $i$, the first-order displacement potential supplied to MUSIC is
\begin{equation}
T_{\delta,i}^{\rm MUSIC}(k)
=
-\frac{d_i(k)}{k_{\rm Mpc}^2},
\qquad
k_{\rm Mpc}=h\,k,
\label{eq:density_music}
\end{equation}
with $k$ tabulated in $h\,\mathrm{Mpc}^{-1}$.

CLASS is run in synchronous gauge. The velocity divergence is transformed
explicitly to Newtonian gauge using
\begin{equation}
\alpha
=
\frac{h_{\rm syn}'+6\eta_{\rm syn}'}{2k_{\rm Mpc}^2},
\qquad
\theta_i^{\rm N}
=
\theta_i^{\rm S}
+
k_{\rm Mpc}^2\alpha,
\end{equation}
and the scalar velocity transfer used by the MUSIC interface is
\begin{equation}
T_{v,i}^{\rm MUSIC}(k)
=
\frac{\theta_i^{\rm N}(k)}
     {aH\,k_{\rm Mpc}^2}.
\label{eq:velocity_music}
\end{equation}
This distinction follows standard gauge-transformation logic
\citep{MaBertschinger1995} and is important in multi-component initial
conditions \citep{ValkenburgVillaescusa2017}.

As an internal numerical consistency check, we separately reconstruct the Newtonian-gauge SFDM
velocity transfer from the native CLASS velocity-divergence and metric transfer functions
at $z=120$. The reconstructed transfer is compared with the archived MUSIC
input table over $0.2\leq k<30\,h\,\mathrm{Mpc}^{-1}$.
Across the tested wavenumber bands, the median absolute fractional
discrepancies range from $0.00034\%$ to $0.01217\%$,
while the maximum discrepancy is below $0.094\%$.
The signed CLASS transfer is interpolated linearly
in $\ln k$ onto the MUSIC wavenumber grid. Within each tested band, modes with transfer
amplitudes below $1\%$ of the maximum absolute CLASS transfer amplitude in that band are excluded to avoid unstable relative errors near zero crossings.

The agreement demonstrates internal consistency between the
CLASS gauge-converted velocity transfer and the corresponding
MUSIC input table. An independent cross-code validation of the
underlying synchronous-gauge SFDM velocity divergence is presented
in Sec.~\ref{sec:independent_axiecamb}, yielding a normalized
RMS discrepancy of $0.328\%$ over
$0.20944\leq k\leq13.4\,h\,\mathrm{Mpc}^{-1}$,
without independent velocity-amplitude fitting.
Together, these complementary tests support the consistency
of the SFDM velocity transfer and its conversion to the
MUSIC input convention.

For the box modes used here, the corresponding density gauge correction is
numerically negligible near the scalar cutoff: for the SFDM density the
median absolute synchronous-to-Newtonian fractional correction is
$1.74\times10^{-4}\%$ and the 95th percentile is $5.40\times10^{-3}\%$.
We therefore retain the native synchronous density while transforming the
velocity explicitly. This is an empirically tested approximation for this
setup, not a general statement that the gauges are interchangeable.

\subsection{MUSIC-to-AxioNyx wave construction and field products}
\label{subsec:music_axionyx}

The matched MUSIC outputs are passed to AxioNyx v1.1.6, branch
\texttt{mixed\_dm}, commit \texttt{1aba393}, using the
\texttt{Exec/MDM\_NoHydro} configuration
\citep{HahnAbel2011,Schwabe2020}.  We use the AxioNyx stage to test
the conversion from the realized density and velocity fields to the wave
variables required by the Schr\"odinger--Poisson solver.  In the notation of
Section~\ref{sec:model}, the initialized amplitude is fixed by
$|\Psi|=\sqrt{\rho_{\phi,c}}$, while the phase is reconstructed from the
MUSIC-generated velocity field through the gradient relation in
equation~(\ref{eq:vphase}).  The AxioNyx run log explicitly identifies this
path as a MUSIC-generated axion phase.

For the high-resolution comparison we generated common-parent $512^3$ and $1024^3$
MUSIC realizations in the same $30\,\mpch$ box, with seed 12345 and
$z_{\rm ini}=120$, and initialized corresponding AxioNyx runs
\citep{HahnAbel2011,Schwabe2020}.  The analysis
reported in this paper uses the level-0 initialization output
\texttt{plt00000}; it does not use an evolved nonlinear snapshot.  A one-step
$512^3$ startup run was also completed as an implementation smoke test and
wrote \texttt{plt00001}, but no physical inference is drawn from that evolved
step.

To make the interface reproducible, central planes were exported to CSV at both
resolutions.  The MUSIC products contain the SFDM density and three velocity
components and, separately, the particle/baryonic density and three velocity
components.  The AxioNyx products contain \texttt{AxDens}, \texttt{AxRe},
\texttt{AxIm}, \texttt{AxPhas}, particle mass density, and the three particle
velocity components.  The accompanying \texttt{slice\_statistics.csv} records
the source, resolution, field name, extrema, mean, standard deviation, CSV
filename, and original data path for each exported plane.  
As an additional particle-velocity consistency check, we compare
the AxioNyx particle velocity fields with the corresponding MUSIC
baryon velocity fields using the exported central planes at
$512^3$ and $1024^3$ resolution.

The archived AxioNyx MUSIC initialization code defines the
velocity conversion factor as
$\mathrm{vel\_fac}=L_{\mathrm{Mpc}}h=30$
for the adopted $30\,h^{-1}\,\mathrm{Mpc}$ simulation box.

The component-wise Pearson correlation coefficients are
$0.99969$--$0.99985$ at both resolutions.
The best-fit velocity scaling factors range from
$29.938$ to $30.002$, consistent with the
expected conversion factor.

After fitting a linear scaling and offset, the
normalized RMS residuals range from approximately
$1.73\%$ to $2.50\%$.

These results support the consistency of the particle-velocity
mapping on the exported central planes.

As an end-to-end routing check, the normalized AxioNyx density at
initialization reproduces the MUSIC SFDM density to floating-point precision,
as expected because the wave amplitude is initialized from that density
field.  We therefore treat this comparison as a pipeline-consistency test,
not as an independent convergence measurement.  The non-trivial AxioNyx
checks are instead the phase-link reconstruction and the resolution
sensitivity of the complex field.
\paragraph{Three-dimensional velocity zero-mode validation.}

To examine the velocity zero-mode treatment in the MUSIC-to-AxioNyx
initialization workflow, we evaluated the spatial means of all three
baryonic and dark-matter velocity components over the complete
three-dimensional MUSIC initial-condition domains. For each velocity
component $v_i$, we define the relative zero-mode amplitude as
\begin{equation}
R_i =
\frac{|\langle v_i\rangle|}{\sigma(v_i)},
\qquad i \in \{x,y,z\},
\label{eq:velocity_zero_mode}
\end{equation}
where $\langle v_i\rangle$ is the volume-averaged velocity and
$\sigma(v_i)$ is its standard deviation.

For the $64^3$ and $256^3$ realizations, the largest relative
zero-mode amplitudes among the six baryonic and dark-matter velocity
components were $1.09\times10^{-17}$ and $8.87\times10^{-17}$,
respectively. Both realizations therefore satisfy the adopted
diagnostic criterion $R_i<10^{-6}$, with residual spatial means
consistent with floating-point roundoff.

We also inspected the archived AxioNyx initialization source,
based on commit \texttt{1aba393c407220682bdaaa4e63dddd53ce568963}.
The source reconstructs the wavefunction phase from the velocity
divergence using a multilevel Poisson solve. Since a spatially
uniform velocity component has vanishing divergence, this
construction is insensitive to such a component. However, no
explicit subtraction of the velocity-component means was
identified in the inspected initialization branches.

These numerical tests verify negligible global velocity offsets
in the examined $64^3$ and $256^3$ MUSIC outputs. They do not
establish the corresponding zero-mode properties of the separate
$512^3$ and $1024^3$ production realizations or independently
validate the reconstructed wavefunction current.

\subsection{Independent AxiECAMB validation and numerical precision}
\label{sec:independent_axiecamb}

We performed an independent comparison of the linear SFDM perturbations
computed with CLASS and AxiECAMB \citep{LiuHuGrin2025}, using the
fiducial scalar mass
$m_\phi=3.1623\times10^{-22}\,\mathrm{eV}$ at the initialization
redshift $z_{\rm ini}=120$. The background cosmology and dark-sector
abundances were matched between the implementations, including the
residual CDM component used in the CLASS calculation. The comparison
employed the public AxiECAMB source at Git revision
\texttt{b7ca9ba80aca178da5003d864b8e20cb5905555b}, supplemented
with a diagnostic export of the scalar effective-fluid momentum
perturbation.

The density-transfer comparison was performed after accounting for
the different CLASS and AxiECAMB transfer conventions. In particular,
the common relative sign was established from the scalar-density
perturbations before comparing the velocity transfers. We reconstructed
the AxiECAMB synchronous-gauge velocity divergence from its
effective-fluid momentum variable,
\begin{equation}
\theta_\phi(k)
=
\frac{k\,u_\phi(k)}{1+w_\phi},
\label{eq:axiecamb_velocity}
\end{equation}
where $k$ is expressed in $\mathrm{Mpc}^{-1}$ and
$u_\phi=(1+w_\phi)v_\phi$ in the adopted AxiECAMB convention.
No additional velocity-amplitude normalization was fitted.
The dense CLASS transfer functions were evaluated at the AxiECAMB
wavenumbers by interpolation in $\ln k$.

We quantified the cross-code agreement through the normalized
Euclidean residual
\begin{equation}
\epsilon_\theta =
\frac{
\left[
\sum_i
\left(
\theta_{\phi,i}^{\rm AxiECAMB}
-
\theta_{\phi,i}^{\rm CLASS}
\right)^2
\right]^{1/2}
}{
\left[
\sum_i
\left(
\theta_{\phi,i}^{\rm CLASS}
\right)^2
\right]^{1/2}
}.
\label{eq:velocity_crosscode_residual}
\end{equation}
For the AxiECAMB calculation with
\texttt{accuracy\_boost=2} and
\texttt{transfer\_k\_per\_logint=32}, the comparison over
$0.20944\leq k\leq13.4\,\hmpc$ includes 133 sampled wavenumbers
and yields $\epsilon_\theta=0.3277\%$, with a cosine similarity
of $0.99999922$. Over the cutoff-region interval
$8\leq k\leq13.4\,\hmpc$, the velocity residual is
$0.3320\%$, while the corresponding scalar-density residual
is $0.3666\%$. At larger scales,
$0.05\leq k\leq1\,\hmpc$, the density and velocity residuals
are $0.0101\%$ and $1.6834\%$, respectively.

We further examined the sensitivity of the velocity comparison
to the transfer-function sampling. At fixed
\texttt{accuracy\_boost=2}, increasing
\texttt{transfer\_k\_per\_logint} from 8 to 32 changed the
normalized velocity residual over
$0.20944\leq k\leq13.4\,\hmpc$ from $0.3578\%$ to
$0.3277\%$. Because the two calculations use different native
wavenumber grids, this comparison is interpreted as a sampling
stability check rather than a formal pointwise convergence test.
The results establish close cross-code consistency of the linear
synchronous-gauge scalar perturbations over the tested Fourier
range. They do not independently validate the subsequent
gauge conversion required by the MUSIC interface or the
velocity field reconstructed from the initialized AxioNyx
wavefunction.

As a separate numerical check, we examined the sensitivity of
the CLASS scalar-component suppression scales to increased
integration precision. The fiducial SFDM and matched
$\Lambda$CDM dense800 calculations were repeated using the
high-precision settings in \texttt{pk\_ref.pre}. For the
oscillatory scalar-field perturbations, the explicit
\texttt{evolver=0} setting was removed, retaining the stiff
\texttt{ndf15} integrator.

The suppression scales were determined from the first downward
crossings of the scalar-component power ratio defined in
Section~\ref{sec:cutoff}. The resulting values are summarized
in Table~\ref{tab:precision}. In particular, the half-power
position changes from $k_{50}^{\phi}=10.535589$ to
$10.535142\,\hmpc$, corresponding to a fractional shift
of $-0.00425\%$. This small change demonstrates the stability
of the CLASS half-power extraction under the tested precision
settings, without establishing an equivalent absolute accuracy
for the independent cross-code comparison.
The remaining suppression scales exhibit similarly small
changes under the increased precision settings.
Specifically, the fractional shifts in
$k_{90}^{\phi}$, $k_{75}^{\phi}$, and $k_{25}^{\phi}$
are $-0.00089\%$, $-0.00196\%$, and $-0.00331\%$,
respectively. All four crossing scales therefore
exhibit shifts smaller than $0.005\%$ in magnitude,
with a consistent direction of change. These results
quantify the sensitivity of the extracted suppression
scales to the tested CLASS integration settings,
rather than uncertainties associated with the
physical scalar-field model.

\begin{table}
\centering
\caption{\label{tab:precision}
Sensitivity of the scalar-component suppression scales to
increased numerical precision in CLASS. All crossing
wavenumbers are expressed in $h\,\mathrm{Mpc}^{-1}$.
The reported shifts characterize the internal precision
test rather than the absolute cross-code accuracy.}
\begin{tabular}{lrrr}
\toprule
Crossing & Original & High precision & Shift [\%]\\
\midrule
$k_{90}^{\phi}$ & 6.742600 & 6.742540 & -0.00089\\
$k_{75}^{\phi}$ & 8.580576 & 8.580408 & -0.00196\\
$k_{50}^{\phi}$ & 10.535589 & 10.535142 & -0.00425\\
$k_{25}^{\phi}$ & 12.292598 & 12.292191 & -0.00331\\
\bottomrule
\end{tabular}
\end{table}

\section{Linear cutoff and Fourier-space resolution}
\label{sec:cutoff}

For the dense800 transfer tables, define the large-scale-normalized
scalar-component power ratio
\begin{equation}
S_\phi(k)
=
\frac{
\left|d_\phi^{\rm SFDM}(k)/d_{\rm cdm}^{\Lambda{\rm CDM}}(k)\right|^2
}{
\operatorname{median}_{0.01\leq k\leq0.1}
\left[\left|d_\phi^{\rm SFDM}/d_{\rm cdm}^{\Lambda{\rm CDM}}\right|^2\right]
}.
\end{equation}
The corresponding scalar crossings are
\begin{align}
k_{90}^{\phi}&=6.7426\,\hmpc, &
k_{75}^{\phi}&=8.5806\,\hmpc,\\
k_{50}^{\phi}&=10.5356\,\hmpc, &
k_{25}^{\phi}&=12.2926\,\hmpc.
\end{align}
For each requested level, the reported value is the \emph{first downward
crossing encountered as $k$ increases} from the large-scale plateau. This
removes ambiguity if later high-$k$ oscillations generate additional
crossings. These values are generated from the dense800 files
\texttt{sfdm\_m215\_ICdense800\_tk.dat} and
\texttt{lcdm\_z120\_dense800\_tk.dat}, each with 5084 sampled
wavenumbers, using interpolation in $\ln k$ between the samples that bracket
each requested power level.

For a uniform $N^3$ grid in a box of side $L$ we adopt
\begin{equation}
k_{\rm ana}
\equiv
\frac{\kny}{2}
=
\frac{\pi N}{2L}
\label{eq:halfny}
\end{equation}
as a conservative analysis boundary, not as a universal convergence theorem.
For $L=30\,\mpch$, the values are
$3.35$, $6.70$, $13.40$, $26.81$, and $53.62\,\hmpc$ for
$N=64,128,256,512,1024$, respectively.

\begin{figure}
\centering
\includegraphics[width=0.98\columnwidth]{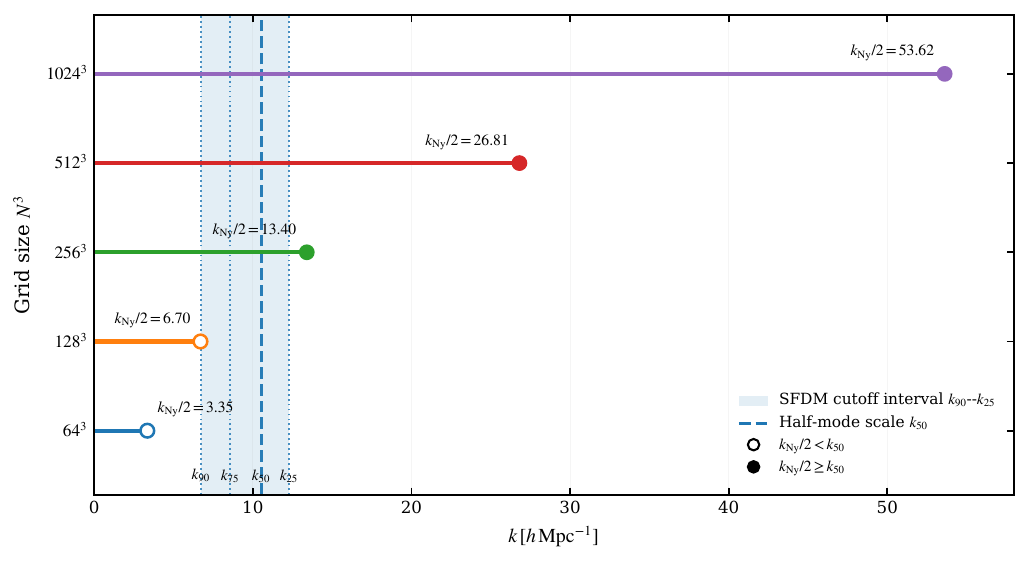}
\caption{Fourier-space resolution reach compared with the dense CLASS-SFDM cutoff at
$z_{\rm ini}=120$ for a $30\,\mpch$ box.  The shaded interval spans the
fiducial scalar suppression region from $k_{90}^{\phi}$ to $k_{25}^{\phi}$,
and the dashed line marks $k_{50}^{\phi}$.  Horizontal markers show the
adopted half-Nyquist analysis reach $k_{\rm Ny}/2$ for each uniform grid.
The $256^3$ grid is the first listed resolution whose conservative Fourier
range extends beyond the scalar half-power scale; this statement concerns
density-mode coverage only and does not certify wave-phase resolution.}
\label{fig:suppression}
\end{figure}

Fig.~\ref{fig:suppression} answers only a Fourier-space question.  It does
not establish that the wave phase is adequately sampled; the latter depends
on equation~(\ref{eq:phase_nyquist}).

\section{Matched initial conditions and density recovery}
\label{sec:icvalidation}


\subsection{MUSIC density recovery}

The accepted MUSIC v1.53 configuration uses first-order Lagrangian
perturbation theory (1LPT; \texttt{use\_2LPT=no} and
\texttt{use\_LLA=no}), $z_{\rm ini}=120$, a $30\,\mpch$ box, common
random phases within matched SFDM--$\Lambda$CDM pairs, and the validated
normalization
\texttt{force\_pnorm=3.1586221390e-11}.
Because the present analysis stops at the initial-condition stage, we do not
use a starting-redshift sequence to assess subsequent nonlinear 1LPT
transients\citep{Crocce2006,Michaux2021}.

The Fourier estimator is applied directly to the raw MUSIC Level-0
\emph{SFDM-component density} grid and compared with the corresponding
CLASS prediction for the \emph{same SFDM component}, rather than with the
CLASS total-matter power or the separate baryonic/particle component. Thus
both numerator and denominator of the reported MUSIC/CLASS power ratio refer
to the SFDM component. This test isolates the supplied density field and the
transfer-function interface rather than particle deposition. No
MUSIC-derived half-power location is compared with a linear half-power
location unless both employ identical component and reference definitions. For
reproducibility, the analysis records the shell-bin edges, mode counts,
random-amplitude prescription, and the script used to construct the statistic.
The quoted RMS therefore measures the discrepancy between the realized field
and the linear target and should not be interpreted solely as a numerical
error estimate.

Across $1\leq k\leq13.4041\,\hmpc$, the corrected $256^3$ realization has
median
$P_{\rm MUSIC}/P_{\rm CLASS}=1.00156$,
mean $0.99960$, and an RMS fractional deviation of $1.70\%$.
Restricting the comparison to the scalar-cutoff interval
$8\leq k\leq13.4041\,\hmpc$ gives a median ratio of $1.00131$ and an RMS
fractional deviation of $0.138\%$.

This result is deliberately interpreted as a density-recovery test. Agreement
of the realized density power with the CLASS target does not, by itself,
demonstrate that the complex wave phase is sufficiently sampled.


\subsection{Component-matched MUSIC--CLASS half-power validation}
\label{subsec:component_matched_halfpower}

The suppression comparison requires matching the numerator and denominator
components in the realized MUSIC and linear CLASS fields; scalar-component
and total-matter power ratios must not be interchanged.

We therefore define the component-matched suppression ratio as

\begin{equation}
\mathcal{S}_{\phi}(k,z)
=
\frac{
P_{\phi}^{\mathrm{SFDM}}(k,z)
}{
P_{c}^{\Lambda\mathrm{CDM}}(k,z)
},
\label{eq:comment10_component_ratio}
\end{equation}

where $P_{\phi}^{\mathrm{SFDM}}$ is the scalar-field density
auto-power spectrum and $P_{c}^{\Lambda\mathrm{CDM}}$ is the
cold-dark-matter density auto-power spectrum in the reference
$\Lambda$CDM cosmology.

Unlike total-matter power, this ratio contains no baryonic or
cross-component contribution.

For the linear calculation, the corresponding component-matched
suppression is obtained directly from the CLASS density transfer
functions,

\begin{equation}
\mathcal{S}_{\phi}^{\mathrm{CLASS}}(k,z)
=
\left|
\frac{
d_{\mathrm{sfdm}_1}^{\mathrm{SFDM}}(k,z)
}{
d_{\mathrm{cdm}}^{\Lambda\mathrm{CDM}}(k,z)
}
\right|^2.
\label{eq:comment10_class_ratio}
\end{equation}

Because the calculations employ a common primordial power-spectrum
prescription, the primordial factor cancels in the ratio.

Following the linear cutoff definition introduced in
Sec.~\ref{sec:cutoff}, the suppression is normalized to its
large-scale value,

\begin{equation}
\widehat{\mathcal{S}}_{\phi}^{\mathrm{CLASS}}(k,z)
=
\frac{
\mathcal{S}_{\phi}^{\mathrm{CLASS}}(k,z)
}{
\displaystyle
\operatorname{median}_{0.01\leq q\leq0.1}
\mathcal{S}_{\phi}^{\mathrm{CLASS}}(q,z)
},
\label{eq:comment10_normalized_class}
\end{equation}

where the normalization interval is expressed in
$h\,\mathrm{Mpc}^{-1}$.

The scalar-component half-power scale is defined by the first
downward crossing satisfying

\begin{equation}
\widehat{\mathcal{S}}_{\phi}
\left(k_{1/2},z\right)
=
\frac{1}{2}.
\label{eq:comment10_halfpower_definition}
\end{equation}

Using the archived dense CLASS transfer tables at $z=120$, the
unbinned linear calculation gives

\begin{equation}
k_{1/2}^{\phi,\mathrm{CLASS}}
=
10.535586\,\hmpc.
\label{eq:comment10_class_halfpower}
\end{equation}

We independently test this scale using the original complete
$256^3$ SFDM and matched $\Lambda$CDM MUSIC initial-condition
realizations. Both realizations use a
$30\,h^{-1}\mathrm{Mpc}$ periodic box, the same initial redshift,
and matched random phases.

The reconstructed fields are obtained from the three-dimensional
dark-matter displacement components stored in the original MUSIC
outputs. Since the simulations employ first-order Lagrangian
perturbation theory, the linear density contrast can be recovered
from the Fourier-space divergence of the displacement field,

\begin{equation}
\delta_{\mathrm{rec}}(\mathbf{k})
=
-i\mathbf{k}\cdot
\boldsymbol{\Psi}(\mathbf{k}),
\label{eq:comment10_density_reconstruction}
\end{equation}

where $\boldsymbol{\Psi}$ denotes the first-order displacement
field in the coordinate convention used for the reconstruction.

The displacement reconstruction uses the Fourier convention identified
in the retained MUSIC configurations and source-code inspection.

For the reconstructed MUSIC fields, we evaluate the
component-matched power suppression as

\begin{equation}
\mathcal{S}_{\phi}^{\mathrm{MUSIC}}(k)
=
\frac{
P_{\phi}^{\mathrm{SFDM,MUSIC}}(k)
}{
P_{c}^{\Lambda\mathrm{CDM},\mathrm{MUSIC}}(k)
}.
\label{eq:comment10_music_ratio}
\end{equation}

The CLASS prediction is evaluated on the same discrete Fourier modes as
each finite-volume MUSIC shell.

For a shell $B$, the mode-matched theoretical suppression is

\begin{equation}
\mathcal{S}_{\phi,B}^{\mathrm{CLASS}}
=
\frac{
\displaystyle
\sum_{\mathbf{k}\in B}
P_{\phi}^{\mathrm{SFDM,CLASS}}(k)
}{
\displaystyle
\sum_{\mathbf{k}\in B}
P_{c}^{\Lambda\mathrm{CDM},\mathrm{CLASS}}(k)
}.
\label{eq:comment10_mode_matched_class}
\end{equation}

The theoretical powers include the common primordial spectral weighting;
we do not substitute a single shell-center transfer amplitude.

For the adopted box size, the fundamental wavenumber is

\begin{equation}
k_{\mathrm{fund}}
=
\frac{2\pi}{L}
=
0.20944\,\hmpc.
\label{eq:comment10_fundamental_mode}
\end{equation}

The large-scale normalization interval used for the unbinned
CLASS transfer functions is therefore inaccessible to the
finite-volume MUSIC realization.

Accordingly, for the MUSIC comparison and its mode-matched CLASS
counterpart, we normalize the suppression over

\begin{equation}
0.3\leq k\leq1.0\,\hmpc.
\label{eq:comment10_music_normalization}
\end{equation}

Both the reconstructed MUSIC suppression and the mode-matched
CLASS suppression are normalized over this same interval
before their half-power crossings are determined.

To assess sensitivity to the Fourier-shell discretization,
we repeat the comparison for five logarithmic bin widths,
$\Delta\ln k=0.05$, $0.08$, $0.10$, $0.15$, and $0.20$.
The results are presented in
Table~\ref{tab:comment10_component_validation}.


\begin{table}[t]
\centering

\caption{
Component-matched MUSIC--CLASS half-power validation at
$z_{\rm ini}=120$ using the complete $256^3$
initial-condition realizations.
The suppression is defined using the SFDM scalar-density
power in the numerator and the matched $\Lambda$CDM
CDM-density power in the denominator.
The CLASS prediction is evaluated over the same discrete
Fourier modes as the MUSIC estimator.
Here,
$\Delta k_{1/2}=
k_{1/2}^{\rm MUSIC}-
k_{1/2}^{\rm CLASS,bin}$.
The last column gives the RMS fractional discrepancy
between the normalized suppression ratios in the cutoff region.
All reported cutoff scales and differences are in
$h\,\mathrm{Mpc}^{-1}$.
}

\label{tab:comment10_component_validation}

\begin{ruledtabular}
\begin{tabular}{ccccc}

$\Delta\ln k$
&
$k_{1/2}^{\rm MUSIC}$
&
$k_{1/2}^{\rm CLASS,bin}$
&
$\Delta k_{1/2}$
&
RMS (\%) \\

\hline

0.05 & 10.53748 & 10.53842 & -0.00095 & 0.018 \\

0.08 & 10.53657 & 10.53600 & +0.00057 & 0.044 \\

0.10 & 10.53827 & 10.53673 & +0.00155 & 0.040 \\

0.15 & 10.53654 & 10.53541 & +0.00113 & 0.030 \\

0.20 & 10.60524 & 10.60402 & +0.00122 & 0.039 \\

\end{tabular}
\end{ruledtabular}

\end{table}


The reconstructed MUSIC suppression agrees closely with the
component-matched CLASS prediction.

For the four narrower binning choices,
$\Delta\ln k\leq0.15$, the measured MUSIC half-power
wavenumbers lie between approximately
$10.5365$ and $10.5383\,\hmpc$,
consistent with the unbinned CLASS scalar-component
half-power scale of $10.5356\,\hmpc$.

For the coarsest tested binning,
$\Delta\ln k=0.20$, both the MUSIC and mode-matched
CLASS half-power estimates shift to approximately
$10.605\,\hmpc$.
The common shift reflects sensitivity of the interpolated
crossing to the adopted Fourier-shell width rather than
a comparable discrepancy between MUSIC and CLASS.

When both calculations are evaluated with identical
Fourier-mode sampling and binning, the absolute difference
between their measured half-power positions satisfies

\begin{equation}
\left|
k_{1/2}^{\mathrm{MUSIC}}
-
k_{1/2}^{\mathrm{CLASS,bin}}
\right|
<
1.6\times10^{-3}\,\hmpc
\label{eq:comment10_halfpower_agreement}
\end{equation}

for all five tested binning choices.

The corresponding RMS fractional discrepancies between
the normalized MUSIC and mode-matched CLASS suppression
ratios in the cutoff region range from
$0.018\%$ to $0.044\%$.

These results provide a direct component-consistency
check of the finite-volume MUSIC initial conditions
against the corresponding linear-theory calculation.
The different binning choices are alternative analyses
of the same initial-condition realizations, rather
than independent statistical measurements.

As an additional transfer-interface check, the archived
SFDM CAMB-format density transfer was compared with
the original CLASS scalar-density transfer function,
using the documented CLASS-to-CAMB density convention,

\begin{equation}
T_{\delta,i}^{\mathrm{CAMB}}(k)
=
-\frac{d_i^{\mathrm{CLASS}}(k)}
       {k_{\mathrm{Mpc}}^2},
\qquad
k_{\mathrm{Mpc}}=hk.
\label{eq:comment10_class_camb_conversion}
\end{equation}

The archived $\Lambda$CDM CLASS configuration and transfer
functions were also checked against the original MUSIC
cosmological parameters. A separate reconstructed
$\Lambda$CDM density-transfer diagnostic was produced
from the preserved CLASS table.

However, the original historical $\Lambda$CDM
CAMB-format input file used by MUSIC was not recovered.
Consequently, its exact byte-level provenance cannot
be independently established from the retained files.
The reconstructed density-transfer diagnostic is
not treated as the historical MUSIC input.

The agreement reported in
Table~\ref{tab:comment10_component_validation}
is instead based on the original retained MUSIC
displacement fields and the independently evaluated
component-matched CLASS prediction.

The results establish numerical consistency of
the component-matched scalar-density suppression
at the initial-condition stage.
They do not constitute a test of nonlinear evolved
power spectra or demonstrate convergence of the
complex SFDM wave phase.

\subsection{Common-parent provenance and shared-mode verification}

We next compare matched $512^3$ and $1024^3$ initial conditions generated as
a common-parent resolution pair in the same $30\,\mpch$ box. The retained
MUSIC products are identified in the generation record as
\texttt{IC\_SFDM\_WAVE\_COMMONPARENT512} and
\texttt{IC\_SFDM\_WAVE\_COMMONPARENT1024}. The corresponding AxioNyx
products are taken from the associated common-parent initialization runs.

As an independent check that the retained products trace the same large-scale
realization, we block-average the native $1024^2$ central SFDM-density slice
to the effective $512^2$ sampling and compare it with the native $512^2$
slice in both real and Fourier space. The resulting real-space Pearson
correlation is
\begin{equation}
r_\delta=0.998910.
\end{equation}
Over the conservative shared Fourier range,
\begin{equation}
k\leq \frac{k_{\rm Ny}^{512}}{2}
      =26.8083\,\hmpc,
\end{equation}
the complex Fourier coherence is $0.998910$, with an aggregate cross-phase
consistent with zero. On the largest scales,
$0.209<k<2\,\hmpc$, the coherence increases to $0.999611$, while over
$2<k<8\,\hmpc$ and the scalar-cutoff interval
$8<k<13.4\,\hmpc$ the corresponding coherences are $0.998484$ and
$0.998161$, respectively.

These diagnostics provide strong numerical evidence that the retained
central-plane products trace the same large-scale parent realization rather
than two independent random fields. They do not constitute an
element-by-element proof that every overlapping three-dimensional Fourier
coefficient is identical, because the original full three-dimensional
parent-noise fields are not part of the retained analysis products. We
therefore interpret the $512^3$--$1024^3$ comparison as a common-parent
resolution-sensitivity test rather than as a strict fixed-field convergence
sequence.

The associated phase diagnostics were generated with AxioNyx v1.1.6, branch
\texttt{mixed\_dm}, commit \texttt{1aba393}. Only the retained central-plane
products are used in the comparison. The fine realization contains additional
high-$k$ structure, so differences between the two resolutions are expected
even when their large-scale parent realization is shared.

\subsection{AxioNyx density and phase comparison}

We compare the AxioNyx wave fields constructed from the matched
common-parent $512^3$ and $1024^3$ MUSIC realizations. The analysis
uses the retained Level-0 central-plane products from the uniform-grid
calculations. No coarse--fine AMR initialization is evaluated here.
Because the $1024^3$ realization contains additional high-$k$
structure, this comparison is interpreted as a resolution-sensitivity
test rather than strict fixed-band convergence.

For the central SFDM density plane, the $1024^3$ field is averaged
in $2\times2$ blocks before comparison with the native $512^2$
plane. The Pearson correlation,
$r_\delta=0.998910$, was established in the shared-mode verification
above. The RMS density residual is
$2.3357\times10^{-3}$, corresponding to $4.669\%$ of the standard
deviation of the $512^3$ central slice. Thus, the density morphology
remains strongly correlated after matching the effective spatial
sampling.

For the phase comparison, we reconstruct the wrapped wavefunction
phase directly from the real and imaginary AxioNyx fields:
\begin{equation}
\theta_N(\bm{x})
=
\arg\!\left[
\mathrm{AxRe}_N(\bm{x})
+i\,\mathrm{AxIm}_N(\bm{x})
\right].
\end{equation}
The corresponding unit-complex phase field is defined as
\begin{equation}
u_N(\bm{x})
=
\frac{\Psi_N(\bm{x})}{|\Psi_N(\bm{x})|}
=
e^{i\theta_N(\bm{x})}.
\end{equation}
This definition is used consistently for the phase-comparison
statistics reported below.

The native $1024^2$ phase field is coarse-grained circularly,
rather than by directly averaging phase angles. Specifically,
we define
\begin{equation}
Z
=
\left\langle u_{1024}\right\rangle_{2\times2},
\qquad
R=|Z|,
\qquad
u_{1024,b}
=
\frac{Z}{R},
\end{equation}
where $R$ measures the local coherence of the fine-grid phases
within each $2\times2$ block. The normalized coarse-grained phase
is defined for $R>0$.

The wrapped phase difference between the coarse- and fine-grid
representations is then
\begin{equation}
\Delta\theta
=
\arg\!\left(
u_{1024,b}u_{512}^{\ast}
\right),
\end{equation}
with $\Delta\theta\in[-\pi,\pi]$.

We first apply the coherence threshold $R\geq0.5$.
This selection retains $227006$ of the $262144$ central-plane
blocks, corresponding to $86.596\%$ of the plane. The remaining
$35138$ blocks ($13.404\%$) are excluded from the masked
statistics.

For the retained blocks, we obtain
\begin{align}
{\rm median}\,|\Delta\theta|
&=0.246443~{\rm rad},\\
P_{90}(|\Delta\theta|)
&=0.753380~{\rm rad},\\
\langle\cos\Delta\theta\rangle
&=0.911654,\\
\left|
\left\langle e^{i\Delta\theta}\right\rangle
\right|
&=0.934456.
\end{align}
The final quantity represents the magnitude of the mean
unit-complex phase difference evaluated over the retained
$R\geq0.5$ blocks.

To quantify the sensitivity of these results to the coherence
selection, we repeat the analysis without imposing the
$R\geq0.5$ threshold. All $262144$ central-plane blocks have
defined coarse-grained phases in the present dataset
($R>0$), allowing the full plane to be included.

The unmasked statistics are
\begin{align}
{\rm median}\,|\Delta\theta|
&=0.263796~{\rm rad},\\
P_{90}(|\Delta\theta|)
&=0.825540~{\rm rad},\\
\langle\cos\Delta\theta\rangle
&=0.895724,\\
\left|
\left\langle e^{i\Delta\theta}\right\rangle
\right|
&=0.924088.
\end{align}
The unmasked mean absolute phase difference is
$0.357517$ rad, with an RMS phase difference of
$0.480746$ rad. The fraction of blocks satisfying
$|\Delta\theta|>\pi/2$ is $0.3422\%$.

Removing the coherence threshold therefore increases the
median absolute phase difference from $0.246443$ to
$0.263796$ rad and the 90th percentile from $0.753380$
to $0.825540$ rad. The unmasked results show moderately
larger phase discrepancies, while preserving the qualitative
interpretation of the central-plane comparison.

These diagnostics quantify the sensitivity of the measured
phase differences to the coherence selection. In particular,
blocks with low $R$ have less well-defined coarse-grained
phase directions and can be more sensitive to small changes
in the underlying complex field. Neither the masked nor
the unmasked comparison establishes full-volume
wavefunction convergence.

\begin{figure*}[t]
\centering
\includegraphics[width=0.98\textwidth]
{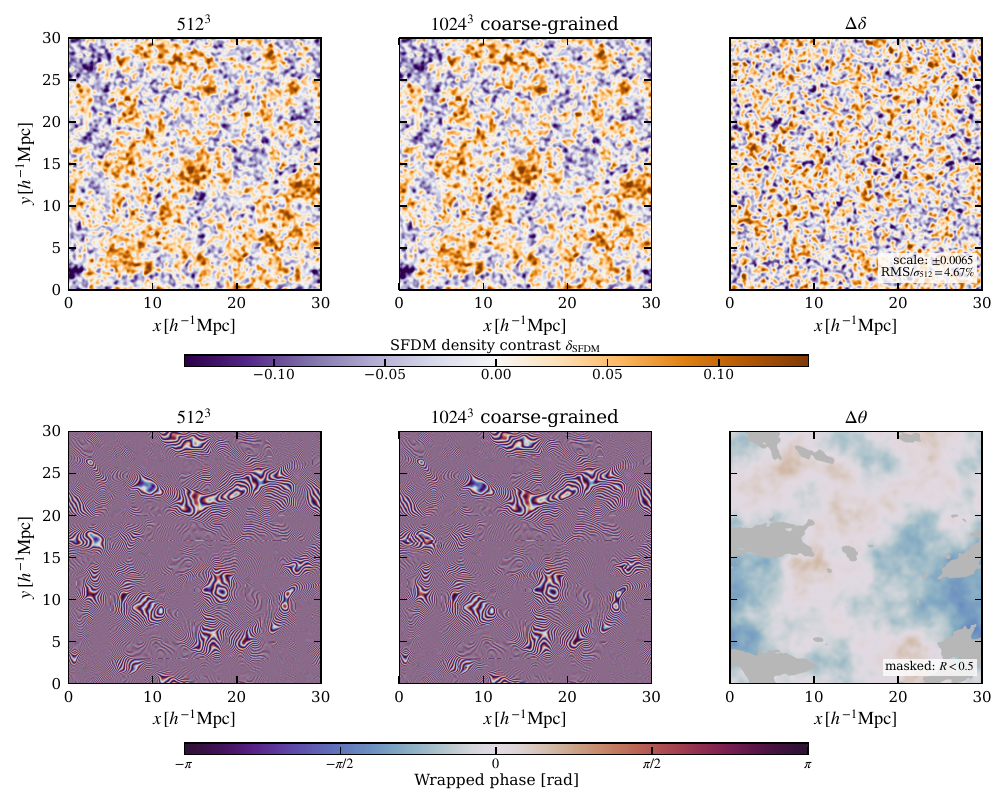}
\caption{
AxioNyx wave-field initialization from the matched common-parent
$512^3$ and $1024^3$ MUSIC realizations at $z_{\rm ini}=120$.
The upper row compares the $512^3$ SFDM density contrast with
the $1024^3$ central plane coarse-grained to the same effective
$512^2$ sampling. The right panel shows the density residual
$\Delta\delta=
\langle\delta_{1024}\rangle_{2\times2}-\delta_{512}$.
The first two density panels share a common color scale,
while the residual uses the smaller symmetric range indicated
in the panel.
The lower row shows the corresponding wrapped wavefunction phases
and their circular difference
$\Delta\theta=\arg(u_{1024,b}u_{512}^{\ast})$.
Gray pixels indicate blocks with local fine-grid phase coherence
$R<0.5$, comprising $13.404\%$ of the central plane.
The remaining $86.596\%$ of blocks are displayed in the phase
residual panel. Both masked and unmasked phase statistics are
reported in the text.
Because refinement introduces additional high-$k$ modes,
the residuals characterize resolution sensitivity rather than
strict fixed-field convergence.
}
\label{fig:highres}
\end{figure*}

The main lesson from Fig.~\ref{fig:highres} is not that the
$1024^3$ wave field is formally converged. Rather, the
density morphology remains strongly correlated between
the two resolutions while the complex phase responds more
visibly to the additional small-scale structure.

The agreement between the masked and unmasked diagnostics
supports this qualitative interpretation without requiring
that low-coherence blocks be discarded from the analysis.
Nevertheless, agreement between two resolutions is not
equivalent to satisfying an absolute phase-sampling
criterion. This motivates the independent phase-sampling
test presented next.

\subsection{Phase sampling and velocity-to-phase validation}

The density-recovery and common-parent comparisons presented
above do not, by themselves, establish that the SFDM wave phase
is adequately sampled. We therefore assess phase resolution
using the velocity-implied phase-Nyquist criterion defined
in equation~(\ref{eq:phase_nyquist}), complemented by
direct diagnostics of the initialized complex wave field.

For the fiducial $256^3$ wave initial condition, the
phase-Nyquist velocity is
\begin{equation}
v_{\rm Ny}^{\rm phase}
=
13.2547\,\kms .
\end{equation}
The corresponding velocity RMS values are
$17.613$, $18.578$, and $17.019\,\kms$
in the three Cartesian directions, respectively.
These values exceed the phase-Nyquist velocity,
indicating a substantial risk of phase undersampling.
Consistently, the full-volume $256^3$ analysis
presented in Table~\ref{tab:phase3d} yields
directional violation fractions of approximately
$47.7\%$, $48.3\%$, and $44.0\%$.
The $256^3$ wave initial condition therefore
exhibits substantial phase-Nyquist violations,
despite accurately reproducing the CLASS density
power spectrum over the scalar-cutoff region.

As a complementary diagnostic, we evaluate the maximum
absolute \emph{wrapped} phase difference between each
cell and its six nearest neighbors. For the standalone
$64^3$, $128^3$, and $256^3$ wave fields, the mean
maximum phase differences are $2.69225$, $2.63382$,
and $2.47276$ rad, respectively, with corresponding
medians of $2.79743$, $2.77715$, and $2.63156$ rad.
The fractions of cells with a maximum wrapped
neighbor difference exceeding $\pi/2$ are
$98.4268\%$, $96.0271\%$, and $90.5048\%$.
Because these three realizations do not form a
common-parent hierarchy, the observed trend is
interpreted as a phase-sampling diagnostic rather
than a formal resolution-convergence test.

A separate root-only $128^3$ calculation provides
a direct validation of the MUSIC-to-AxioNyx
velocity-to-phase convention. Comparing the
reconstructed phase-link increments with those
predicted from the input velocity field yields
component-wise RMS ratios of
\begin{equation}
(1.000045,\;1.000038,\;0.999976)
\end{equation}
and Pearson correlation coefficients of
\begin{equation}
(0.999979,\;0.999958,\;0.999952)
\end{equation}
in the $x$, $y$, and $z$ directions, respectively.
The near-unity ratios and high correlations
demonstrate numerical consistency of the implemented
velocity-to-phase mapping at initialization.
However, this test does not independently validate
the underlying CLASS-to-MUSIC SFDM velocity transfer
or establish that the discretized wavefunction current
reproduces the target momentum density,
$\bm j_{\phi,c}=\rho_{\phi,c}\bm v_{\rm target}$.

We further examine phase-resolution scaling using
a controlled common-parent $128^3\rightarrow256^3$
comparison. Unlike the wrapped phase differences
considered above, this test uses the \emph{unwrapped},
velocity-predicted nearest-neighbor phase increments
$\Delta\theta^{(v)}$ defined in
equation~(\ref{eq:dtheta}).
The mean, median, and 90th percentile of the
maximum six-link $|\Delta\theta^{(v)}|$ decrease from
\begin{equation}
(14.1215,\;14.3805,\;18.4420)\ {\rm rad}
\end{equation}
at $128^3$ to
\begin{equation}
(6.9350,\;7.05205,\;9.15795)\ {\rm rad}
\end{equation}
at $256^3$.
The corresponding fine-to-coarse ratios are
\begin{equation}
(0.4911,\;0.4904,\;0.4966),
\end{equation}
consistent with the expected scaling
$\Delta\theta^{(v)}\propto\Delta x$.
Nevertheless, $97.9493\%$ of the refined cells
still exhibit at least one unwrapped,
velocity-predicted nearest-neighbor phase increment
exceeding $\pi$.
This maximum-six-link violation fraction is
distinct from the individual component-wise
fractions reported above and therefore measures
a different aspect of phase undersampling.

At higher resolutions, the full-volume analysis
of the MUSIC velocity fields shows a substantial
improvement in phase sampling. The combined
component-wise phase-Nyquist violation fraction
decreases from $13.7299\%$ at $512^3$ to
$0.124983\%$ at $1024^3$, corresponding to an
approximately $110$-fold reduction.
The remaining violations at $1024^3$ indicate
that phase sampling is substantially improved,
although not completely resolved everywhere.

Taken together, these results demonstrate that
accurate recovery of the linear density power
spectrum is necessary but insufficient to
establish adequate wave-phase resolution.
The low-resolution diagnostics reveal significant
phase undersampling, while the full-volume
high-resolution calculations show a pronounced
reduction in velocity-implied phase-Nyquist
violations. These findings motivate the full-volume phase-resolution
analysis presented in the following subsection and the
independent-realization comparison that follows.

\subsection{Three-dimensional phase-resolution validation}
\label{sec:phase3d}

We extend the phase-sampling analysis to the full
three-dimensional initial-condition volumes at
resolutions of $256^3$, $512^3$, and $1024^3$.
The diagnostic evaluates the unwrapped phase
increments implied by the initial velocity field,
as introduced in Eq.~(\ref{eq:dtheta}).
This quantity is distinct from the wrapped circular
phase differences used in the common-parent
wavefunction comparison.

For each Cartesian direction, we calculate the
fraction of evaluated phase increments satisfying
$|\Delta_i\theta^{(v)}|>\pi$.
The combined violation fraction is the arithmetic
mean of the three directional fractions.
Table~\ref{tab:phase3d} summarizes the
full-volume results.

\begin{table}[t]
\centering
\caption{
Full-volume velocity-implied phase-Nyquist
violation fractions at $z_{\rm ini}=120$.
The combined fraction is the mean of the
three directional fractions.
}
\label{tab:phase3d}
\begin{tabular}{lrrrr}
\toprule
Resolution & $x$ (\%) & $y$ (\%)
& $z$ (\%) & Combined (\%) \\
\midrule
$256^3$  & 47.691 & 48.327 & 44.017 & 46.678 \\
$512^3$  & 13.160 & 14.921 & 12.974 & 13.685 \\
$1024^3$ & 0.072507 & 0.275645 & 0.024028 & 0.124060 \\
\bottomrule
\end{tabular}
\end{table}

The results demonstrate a strong dependence of
phase sampling on spatial resolution.
At $256^3$, approximately $46.7\%$ of the
evaluated directional phase increments exceed
the phase-Nyquist threshold, despite the
accurate recovery of the linear density
power spectrum.
This fraction decreases to $13.685\%$
at $512^3$ and $0.124060\%$ at $1024^3$,
corresponding to an approximately $376$-fold
reduction between the lowest and highest
resolutions.

The directional phase-increment standard
deviations also decrease systematically
with increasing resolution, from approximately
$4.2$ rad at $256^3$ to $2.1$ rad at
$512^3$ and $1.05$ rad at $1024^3$.
This behavior is consistent with the
expected dependence
\begin{equation}
\sigma_{\Delta\theta}
\simeq
\frac{a m_\phi\Delta x}{\hbar}\,
\sigma_v,
\end{equation}
for comparable velocity dispersions,
and illustrates the improvement in
phase-gradient sampling as the grid spacing
decreases.

\begin{figure}[t]
\centering
\includegraphics[width=0.78\columnwidth]
{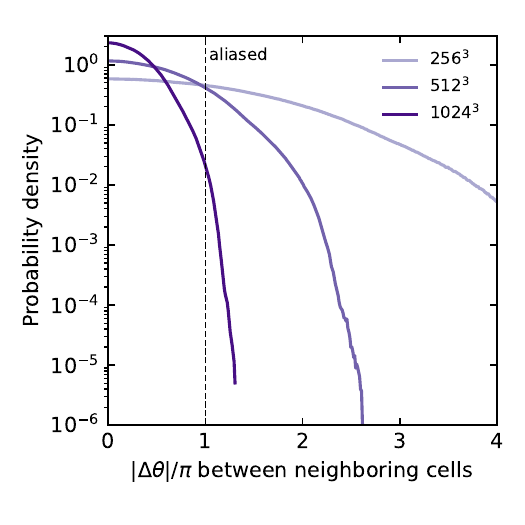}
\caption{
Full-volume distributions of absolute
phase increments at initialization for
the $256^3$, $512^3$, and $1024^3$
realizations, pooled over the three
Cartesian directions.
The dotted vertical line marks the
phase-Nyquist threshold
$|\Delta\theta|=\pi$.
The distributions illustrate the
progressive improvement in phase
sampling with increasing resolution.
}
\label{fig:phase3d_hist}
\end{figure}

Figure~\ref{fig:phase3d_hist} shows the
distributions of absolute phase increments.
The fraction of increments exceeding $\pi$
decreases substantially with increasing
resolution, indicating that the phase
gradients implied by the initial velocity
field are increasingly well sampled.

\begin{figure}[t]
\centering
\includegraphics[width=0.98\columnwidth]
{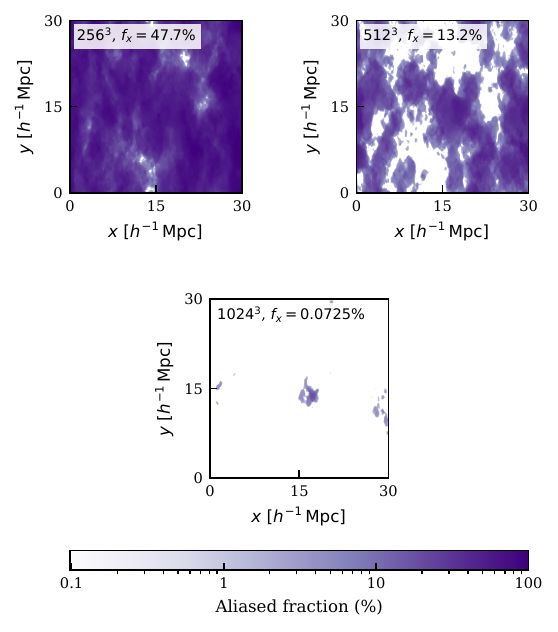}
\caption{
Spatial distribution of phase-Nyquist
violations at $z_{\rm ini}=120$ for
the $256^3$, $512^3$, and $1024^3$
initial conditions.
Each panel shows the fraction of
$x$-direction phase increments
exceeding $\pi$, projected along
the $z$ direction.
A common logarithmic color scale is
used, with white indicating no
detected violations.
The corresponding directional
fractions are $47.691\%$,
$13.160\%$, and $0.072507\%$.
}
\label{fig:phase3d_maps}
\end{figure}

The spatial distribution of phase-Nyquist
violations is presented in
Fig.~\ref{fig:phase3d_maps}.
At $256^3$, violations are widespread
throughout the simulation volume.
Their spatial extent decreases
substantially at $512^3$, while the
remaining violations at $1024^3$
are concentrated in localized regions.

These full-volume diagnostics demonstrate
a substantial improvement in phase sampling
at $1024^3$ for the fiducial scalar mass,
initialization redshift, and box size.
Nevertheless, the remaining violations
show that the phase-Nyquist criterion is
not satisfied everywhere.

Furthermore, these results quantify the
sampling of velocity-implied phase gradients
at initialization. They do not independently
establish convergence of the reconstructed
three-dimensional wavefunction, its
discretized current, or subsequent nonlinear
evolution.
\subsection{Consistency across independent realizations}
\label{subsec:independent_phase}

To assess the consistency of the phase-resolution results,
we compare the full-volume phase-Nyquist statistics from
the reference realization with those obtained from an
additional MUSIC realization at $512^3$ and $1024^3$.
Both calculations use the same fiducial scalar mass,
simulation box size, initialization redshift, and
spatial resolutions.

For the additional MUSIC realization, we estimate
the local velocity-implied phase increment from the
cell-centered velocity components,
\begin{equation}
\widehat{\Delta\theta}_{i}^{(v)}(\mathbf{x})
=
\frac{a m_\phi\Delta x}{\hbar}
v_i(\mathbf{x}),
\qquad i\in\{x,y,z\}.
\label{eq:cell_phase_increment}
\end{equation}
The directional violation fraction is defined as
\begin{equation}
f_i =
\frac{1}{N^3}
\sum_{\mathbf{x}}
\mathbb{I}
\left(
|\widehat{\Delta\theta}_{i}^{(v)}(\mathbf{x})|
>\pi
\right),
\label{eq:directional_phase_fraction}
\end{equation}
where $\mathbb{I}$ denotes the indicator function.
The combined fraction is
\begin{equation}
f_{\rm combined}
=
\frac{f_x+f_y+f_z}{3}.
\label{eq:combined_phase_fraction}
\end{equation}
Here, realization A denotes the original full-volume
phase-sampling analysis, while realization B denotes
the additional analysis based on cell-centered MUSIC
velocity components.

Table~\ref{tab:independent_phase} compares these
measurements with the reference full-volume results
presented in Table~\ref{tab:phase3d}.


\begin{table}[!htbp]
\centering
\caption{
Comparison of full-volume velocity-implied phase-Nyquist
violation fractions for two independently generated SFDM
initial-condition realizations (A and B) using different
random seeds at $z_{\rm ini}=120$.
Both realizations share identical cosmological parameters,
simulation box size, and spatial resolutions.
The combined fraction represents the mean of the three
directional fractions, while differences are expressed
in percentage points (pp).
Both realizations exhibit consistent phase-sampling
statistics and a substantial reduction in phase-Nyquist
violations at $1024^3$, indicating improved phase sampling
at initialization.
}
\label{tab:independent_phase}
\vspace{0.3cm} 
\small
\setlength{\tabcolsep}{5pt}
\renewcommand{\arraystretch}{1.12}

\begin{tabular}{lrrr}
\toprule
Direction &
A (\%) &
B (\%) &
$\Delta$ (pp) \\
\midrule

\multicolumn{4}{c}{$\boldsymbol{512^3}$} \\
\midrule
$x$        & 13.1600 & 13.2145 & 0.0545 \\
$y$        & 14.9210 & 14.9663 & 0.0453 \\
$z$        & 12.9740 & 13.0089 & 0.0349 \\
\textbf{Combined}
           & \textbf{13.6850}
           & \textbf{13.7299}
           & \textbf{0.0449} \\

\midrule
\multicolumn{4}{c}{$\boldsymbol{1024^3}$} \\
\midrule
$x$        & 0.072507 & 0.072918 & 0.000411 \\
$y$        & 0.275645 & 0.277319 & 0.001674 \\
$z$        & 0.024028 & 0.024712 & 0.000684 \\
\textbf{Combined}
           & \textbf{0.124060}
           & \textbf{0.124983}
           & \textbf{0.000923} \\

\bottomrule
\end{tabular}

\end{table}

The two calculations yield closely agreeing
phase-Nyquist violation fractions.
At $512^3$, the combined fractions are
$13.6850\%$ and $13.7299\%$, corresponding
to a relative difference of approximately
$0.33\%$.
At $1024^3$, the corresponding fractions
decrease to $0.124060\%$ and $0.124983\%$,
with a relative difference of approximately
$0.74\%$.

Both calculations exhibit an approximately
$110$-fold reduction in the combined violation
fraction between $512^3$ and $1024^3$.
The close agreement in the directional and
combined statistics supports the consistency
of the observed improvement in phase sampling
across the tested initial conditions.

The additional MUSIC diagnostic uses
cell-centered velocity components to estimate
the local phase increments. Its numerical
agreement with the reference results does
not, by itself, establish equivalence with
a neighboring-link phase-increment estimator.
Consequently, the small differences between
the reported fractions should not be
interpreted solely as random-realization
effects or as a numerical error estimate.

Furthermore, this comparison is restricted
to the initialization stage and does not
establish convergence of the reconstructed
complex wavefunction, its discretized current,
or subsequent nonlinear evolution.
Nevertheless, the agreement provides an
additional consistency check on the
resolution dependence of the phase-Nyquist
diagnostic.
\section{Discussion}
\label{sec:discussion}

The matched initial conditions examined in this work highlight several
numerical requirements that must be assessed independently. The dense CLASS
calculation determines the location and shape of the linear SFDM cutoff,
while the corrected MUSIC realization demonstrates accurate recovery of
the corresponding density power spectrum over the adopted Fourier interval.
However, reproducing the linear density spectrum does not guarantee that
the complex wavefunction can adequately represent the intended velocity
field. The phase-Nyquist criterion provides an additional sampling
requirement, which is substantially violated by the fiducial $256^3$
initial condition despite its accurate density recovery.

The high-precision CLASS calculations establish the numerical
sensitivity of the scalar-component cutoff measurement.
Increasing the CLASS precision changes the half-power
wavenumber $k_{50}^{\phi}$ by approximately $0.00425\%$.
Separately, the independent AxiECAMB comparison yields
a normalized velocity-divergence residual of $0.3277\%$
over the tested Fourier interval. These diagnostics
measure different aspects of numerical consistency
and should not be interpreted as equivalent
uncertainties in the cutoff location.

The common-parent $512^3$--$1024^3$ comparison provides complementary
information about resolution sensitivity. The strong correlation
between the density fields indicates that their large-scale structures
remain consistent after matching the spatial sampling. However,
increasing the resolution introduces additional high-wavenumber modes,
and the resulting differences in the complex phase cannot be interpreted
as a strict fixed-field convergence test. The stronger sensitivity of
the phase to refinement reinforces the need for an absolute
phase-sampling criterion independent of density agreement.

The full-volume velocity-implied phase-Nyquist analysis demonstrates
a substantial improvement in phase sampling at higher resolution.
For the primary realization, the combined violation fraction decreases
from $13.7299\%$ at $512^3$ to $0.124983\%$ at $1024^3$.
A separately generated realization yields closely agreeing values
of $13.6850\%$ and $0.124060\%$, respectively.
The relative differences between the two realizations are
approximately $0.33\%$ at $512^3$ and $0.74\%$ at $1024^3$.
Both exhibit an approximately $110$-fold reduction in the combined
violation fraction as the spatial resolution increases.
This agreement indicates that the observed improvement in phase
sampling is consistent across the two tested random realizations,
rather than being specific to a single realization.

Nevertheless, the small residual violation fraction at $1024^3$
should not be interpreted as evidence of complete phase resolution.
The reported fractions quantify the local velocity-implied phase
increments exceeding the Nyquist threshold at initialization.
They do not independently establish convergence of the full
three-dimensional complex wavefunction, global phase consistency,
or the discretized wavefunction current. Furthermore, agreement
between two realizations provides a consistency check rather than
a statistical convergence test over an ensemble of random seeds.

The AxioNyx initialization stage completes the numerical connection
between the MUSIC density and velocity fields and the complex SFDM
representation. The $512^3$ and $1024^3$ calculations construct the
corresponding wave amplitudes and phases and produce the initialized
field diagnostics. The agreement between the normalized MUSIC and
AxioNyx density fields verifies the routing of the wave amplitude,
while the root-only phase-link test supports the implemented
velocity-to-phase conversion. The full-volume phase-Nyquist analysis
then provides an absolute measure of the ability of the spatial
grid to sample the intended phase gradients. Together, these tests
validate complementary aspects of the initialization procedure
without establishing convergence of subsequent nonlinear evolution.

Several limitations remain relevant to the interpretation of
these results. An independent same-model comparison of the SFDM
momentum-transfer variables and an end-to-end validation of the
reconstructed wave current against
$\bm j_{\phi,c}=\rho_{\phi,c}\bm v_{\rm target}$
have not been completed. Global phase closure, periodic winding,
the treatment of coherent velocity zero modes, and transverse
velocity diagnostics remain additional checks for future
production simulations. These limitations do not invalidate
the reported density-recovery and local phase-sampling
measurements, but they restrict the conclusions that can
be drawn about the full wave initialization.

The implications for Lyman-$\alpha$ forest applications are
therefore methodological rather than observational. Existing
forest constraints on ultralight dark matter require modeling
nonlinear gravitational evolution, baryonic hydrodynamics,
the thermal history of the intergalactic medium, pressure
smoothing, and peculiar-velocity effects
\citep{Irsic2017,Armengaud2017,Kobayashi2017,Nori2019,
RogersPeiris2021,Becker2011,Walther2019}.
The present calculations address the consistency and numerical
resolution of the initial conditions preceding such simulations;
they do not provide nonlinear transmitted-flux predictions
or observational constraints on the SFDM particle mass.
Recent mixed-wave studies further motivate careful treatment
of density and velocity evolution in future applications
\citep{Wang2026}.

The principal outcome of this study is therefore a
validated initialization framework in which the linear
transfer functions, component assignments, gauge conventions,
density recovery, velocity-to-phase mapping, and wave-phase
sampling are evaluated through distinct numerical tests.
In particular, the close agreement between the two
high-resolution realizations strengthens the conclusion
that resolving the linear density spectrum alone is
insufficient: adequate sampling of the associated
wave phase is an independent requirement for reliable
SFDM initial conditions.

\section{Conclusions}
\label{sec:conclusions}

We have constructed and validated matched cosmological initial
conditions for scalar-field dark matter (SFDM) and a reference
$\Lambda$CDM cosmology, examining the numerical consistency
of the initialization pipeline from CLASS-SFDM transfer functions
through MUSIC to the complex wave representation in AxioNyx.
Our analysis demonstrates that accurate recovery of the linear
density power spectrum and adequate sampling of the wave phase
are distinct requirements for reliable SFDM initial conditions.

For the fiducial scalar mass
$m_\phi=3.1623\times10^{-22}\,\mathrm{eV}$ at
$z_{\rm ini}=120$, the dense CLASS calculation yields a
scalar-component half-power scale of
$k_{50}^{\phi}\simeq10.54\,h\,\mathrm{Mpc}^{-1}$.
The internal CLASS precision test changes this scale
by approximately $0.00425\%$, while an independent
AxiECAMB comparison yields a normalized velocity-divergence
residual of $0.328\%$.
The corrected $256^3$ MUSIC realization reproduces the
CLASS SFDM density power spectrum over the adopted
cutoff interval, with a median ratio
$P_{\rm MUSIC}/P_{\rm CLASS}=1.00131$
and an RMS fractional deviation of $0.138\%$.

Despite this density agreement, the phase-sampling diagnostics
show that the $256^3$ wave initial condition remains
substantially under-resolved. The phase-Nyquist velocity
is $13.2547\,\mathrm{km\,s^{-1}}$, below the RMS velocities
of the individual Cartesian components. The controlled
common-parent $128^3\rightarrow256^3$ comparison confirms
the expected reduction in velocity-implied phase increments
with increasing resolution, although $97.95\%$ of the
refined cells retain at least one nearest-neighbor phase
increment exceeding the Nyquist threshold.

At higher resolution, the full-volume velocity-implied
phase-Nyquist violation fraction decreases from
$13.7299\%$ at $512^3$ to $0.124983\%$ at $1024^3$.
A separately generated realization gives corresponding
fractions of $13.6850\%$ and $0.124060\%$.
The relative differences between the two realizations
are approximately $0.33\%$ and $0.74\%$, respectively.
Both therefore exhibit an approximately $110$-fold
reduction in the combined violation fraction.
This agreement supports the robustness of the observed
improvement in phase sampling across the two tested
realizations, rather than attributing the trend to a
particular random realization.

The common-parent $512^3$--$1024^3$ comparison further
demonstrates that the initialized density fields remain
strongly correlated after matching their spatial sampling,
with $r_\delta=0.998910$, while the complex phase is
more sensitive to the additional small-scale structure.
Nevertheless, the small phase-Nyquist violation fraction
at $1024^3$ does not establish complete wavefunction
convergence. The present tests concern the initialization
stage and do not independently validate the discretized
wavefunction current or subsequent nonlinear evolution.

The principal conclusion of this work is that reproducing
the target linear density spectrum is necessary but
insufficient for constructing numerically reliable
wave-based cosmological initial conditions. The velocity
convention, density recovery, phase-gradient mapping,
and phase-Nyquist sampling must be assessed independently.
The validated initialization diagnostics and their
resolution dependence provide a quantitative basis
for future high-resolution SFDM simulations, including
applications to the Lyman-$\alpha$ forest. Predictions
for nonlinear structure formation, transmitted-flux
statistics, and observational constraints on the scalar
mass remain outside the scope of the present study.

\appendix

\section{Sensitivity to initial-condition transfer prescriptions}
\label{app:transfer_prescriptions}

We compare three transfer prescriptions against the component-specific
CLASS-SFDM reference at $z_{\rm ini}=120$, for
$\Omega_{\rm SFDM}=0.264$, $\Omega_{\rm cdm}=10^{-4}$, and
$m_{\rm SFDM}=10^{-21.5}\,{\rm eV}$. These diagnostics test the
sensitivity to assumptions about species and redshift dependence.

The reference calculation uses the component-specific
density and velocity-divergence transfer functions
obtained directly from CLASS in synchronous gauge.
We examine the following alternative prescriptions:

\begin{enumerate}

\item \textit{Longitudinal-gauge velocity reconstruction.}
A diagnostic velocity-divergence transfer is constructed
from the synchronous-gauge metric perturbations using
the linear gauge transformation
\begin{equation}
\theta_i^{\rm L}(k)
=
\theta_i^{\rm S}(k)
+
\frac{h_{\rm syn}'(k)+6\eta_{\rm syn}'(k)}{2},
\label{eq:longitudinal_velocity}
\end{equation}
where the prime denotes differentiation with respect
to conformal time. This reconstruction adopts the standard CLASS metric
and velocity-divergence conventions. The underlying
synchronous-gauge SFDM velocity-divergence transfer
has been independently cross-checked against AxiECAMB,
as described in Sec.~\ref{sec:independent_axiecamb}.
The gauge-converted transfer supplied to MUSIC has
also been verified through the internal consistency
test described following Eq.~\eqref{eq:velocity_music}.
The longitudinal-to-synchronous comparison is therefore
interpreted as a gauge-sensitivity diagnostic rather
than a measure of numerical error.

\item \textit{Scale-independent rescaling from $z=0$.}
The component transfer function at the initial redshift
is approximated by
\begin{equation}
T_{X,i}^{\rm res}(k,z_{\rm ini})
=
A_{X,i}\,T_{X,i}(k,0),
\label{eq:transfer_rescaling}
\end{equation}
where
\begin{equation}
A_{X,i}
=
\frac{
T_{X,i}(k_\star,z_{\rm ini})
}{
T_{X,i}(k_\star,0)
},
\qquad
k_\star=0.05\,h\,{\rm Mpc}^{-1}.
\label{eq:transfer_pivot}
\end{equation}
Here $X$ denotes either density contrast or
velocity divergence. The normalization is evaluated
separately for each component and variable.
This prescription tests the assumption that the
transfer-function shape remains unchanged during
the evolution from $z=0$ to $z_{\rm ini}$.

\item \textit{Total-matter transfer approximation.}
The component-specific transfer function is replaced
by the corresponding total-matter transfer function
at the same redshift,
\begin{equation}
T_{X,i}^{\rm approx}(k,z_{\rm ini})
=
T_{X,\rm tot}(k,z_{\rm ini}).
\label{eq:total_transfer_approx}
\end{equation}

\end{enumerate}

For each approximation, we define the absolute
fractional power difference relative to the
component-specific reference transfer function:
\begin{equation}
\epsilon_{X,i}(k)
=
\left|
\frac{
\left|T_{X,i}^{\rm approx}(k)\right|^2
}{
\left|T_{X,i}^{\rm ref}(k)\right|^2
}
-1
\right|.
\label{eq:transfer_prescription_error}
\end{equation}

Because the compared transfer functions share the
same primordial normalization, the primordial power
spectrum cancels in this ratio.

To avoid ill-conditioned ratios near zeros of the
reference transfer function, we exclude points
satisfying
\begin{equation}
|T_{X,i}^{\rm ref}(k)|
\leq
10^{-4}
\max_{k\in\mathcal{K}}
|T_{X,i}^{\rm ref}(k)|,
\label{eq:transfer_zero_mask}
\end{equation}
where $\mathcal{K}$ denotes the plotted wavenumber
interval. This exclusion is a numerical plotting
criterion and does not represent a physical cutoff.

\begin{figure}[!htbp]
\centering
\includegraphics[width=0.98\columnwidth]
{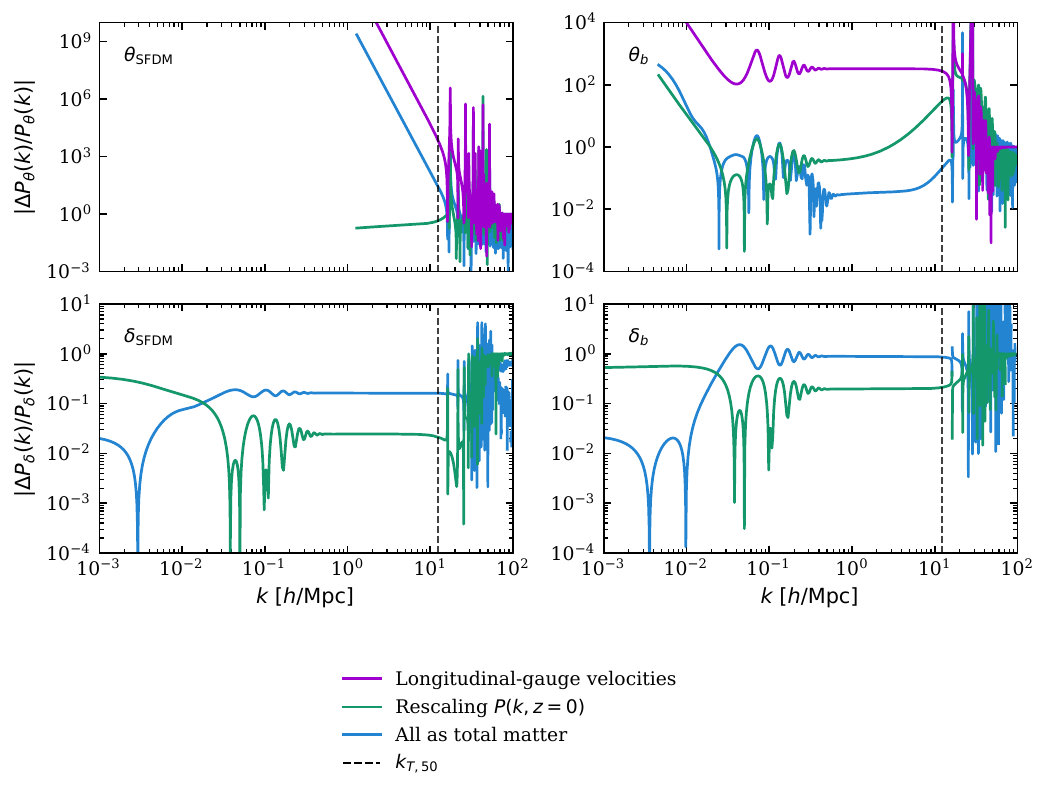}
\caption{
Fractional velocity-divergence (top) and density (bottom)
power differences for SFDM (left) and baryons (right)
at $z_{\rm ini}=120$.
The curves compare three transfer prescriptions with
the synchronous-gauge CLASS reference.
Dashed lines indicate $k=12.29\,\hmpc$,
corresponding approximately to the scalar-component
25\% power-suppression scale $k_{25}^{\phi}$
(equivalently, the 50\% transfer-amplitude threshold).
Reference amplitudes below $10^{-4}$ of their maximum
are excluded.
The differences quantify transfer-prescription sensitivity
rather than numerical initialization errors.
}
\label{fig:sfdm_transfer_prescriptions}
\end{figure}

Figure~\ref{fig:sfdm_transfer_prescriptions}
compares the resulting density and velocity-divergence
power differences for SFDM and baryons.

Large SFDM velocity-panel ratios require particular
caution near zeros of the reference velocity-divergence
transfer function. In these regions, even a small
absolute difference between two transfer prescriptions
can produce a large fractional ratio because the
reference amplitude approaches zero. Consequently,
the apparent enhancement of the fractional difference
does not necessarily imply a correspondingly large
difference in the underlying physical perturbations.
The plotted ratios should therefore be interpreted
as diagnostics of the adopted linear transfer
prescriptions rather than direct measures of numerical
accuracy. In particular, they do not represent
measured errors introduced by MUSIC or AxioNyx,
since the comparison is performed at the level
of the input linear transfer functions, before
the subsequent numerical realization and evolution.

The calculations employ matched CLASS transfer tables
containing 5084 wavenumber samples at both $z=0$
and $z=120$. The newly generated $z=120$ transfers
reproduce the archived reference to the precision
stored in the output files, providing a direct
consistency check of the retained transfer-function
calculation. This agreement establishes reproducibility
of the reference calculation under the tested settings
but does not, by itself, demonstrate numerical
convergence of the small-scale oscillatory features.
Such convergence requires a separate assessment of
the sensitivity to the internal CLASS precision
settings, the sampling of the transfer functions,
and the treatment of rapidly oscillating scalar-field
perturbations. 

\FloatBarrier
\vspace{0.3cm}

\begin{acknowledgments}
HIU and MA gratefully acknowledge support from the SECIHTI PhD
scholarship program. This work received partial funding from
SECIHTI M\'exico under grants CBF-2025-G-1720 and CBF-2025-G-176.
YW acknowledges the computing time granted by the Resource Allocation Board and provided on the supercomputer Emmy at NHR-Nord@Göttingen as part of the NHR infrastructure. Some simulations were conducted with computing resources under the project nip00084.
Additional support was provided by grant I0101/131/07 C-234/07
of the Instituto Avanzado de Cosmolog\'ia (IAC) collaboration
(\url{http://www.iac.edu.mx/}). The authors acknowledge the computational resources provided by the SECIHTI through the Supercomputer Hybrid Cluster
``Xiuhcoatl'' at the General Coordination of Information and
Communication Technologies (CGSTIC) of CINVESTAV
(\url{http://clusterhibrido.cinvestav.mx/}).
\end{acknowledgments}

\section*{Data and Code Availability}

The datasets, simulation inputs, validation records, analysis
scripts, and figures are available on Zenodo\footnote{
\url{https://doi.org/10.5281/zenodo.23719912}}.
The archived code snapshot is identified by commit
\texttt{54d851e}. The archive includes the patched
\texttt{transfer\_camb.cc} for MUSIC v1.53 (revision
\texttt{6248d13}), AxioNyx v1.1.6 (commit \texttt{1aba393}),
and AxiECAMB validation records (revision \texttt{b7ca9ba}).
Full-volume phase diagnostics, reproduction tests, complete
revision identifiers, and SHA-256 checksums are provided
to support reproducibility.

\bibliographystyle{apsrev4-2}
\bibliography{references_validation}

@article{PressRydenSpergel1990,
  author = {Press, William H. and Ryden, Barbara S. and Spergel, David N.},
  title = {Single mechanism for generating large-scale structure and providing dark missing matter},
  journal = {Phys. Rev. Lett.},
  volume = {64},
  pages = {1084--1087},
  year = {1990},
  doi = {10.1103/PhysRevLett.64.1084}
}

@article{Peebles2000,
  author = {Peebles, P. J. E.},
  title = {Fluid Dark Matter},
  journal = {Astrophys. J. Lett.},
  volume = {534},
  pages = {L127--L129},
  year = {2000},
  doi = {10.1086/312677}
}

@article{Hu2000,
  author = {Hu, Wayne and Barkana, Rennan and Gruzinov, Andrei},
  title = {Fuzzy Cold Dark Matter: The Wave Properties of Ultralight Particles},
  journal = {Phys. Rev. Lett.},
  volume = {85},
  pages = {1158--1161},
  year = {2000},
  doi = {10.1103/PhysRevLett.85.1158},
  eprint = {astro-ph/0003365},
  archivePrefix = {arXiv}
}

@article{Hlozek2015,
  author = {Hlo{\v z}ek, Ren{\'e}e and Grin, Daniel and Marsh, David J. E. and Ferreira, Pedro G.},
  title = {A search for ultra-light axions using precision cosmological data},
  journal = {Phys. Rev. D},
  volume = {91},
  pages = {103512},
  year = {2015},
  doi = {10.1103/PhysRevD.91.103512},
  eprint = {1410.2896},
  archivePrefix = {arXiv}
}

@article{UrenaGonzalez2016,
  author = {Ure{\~n}a-L{\'o}pez, L. Arturo and Gonzalez-Morales, Alma X.},
  title = {Towards accurate cosmological predictions for rapidly oscillating scalar fields as dark matter},
  journal = {JCAP},
  volume = {07},
  pages = {048},
  year = {2016},
  doi = {10.1088/1475-7516/2016/07/048},
  eprint = {1511.08195},
  archivePrefix = {arXiv}
}

@article{Hui2017,
  author = {Hui, Lam and Ostriker, Jeremiah P. and Tremaine, Scott and Witten, Edward},
  title = {Ultralight scalars as cosmological dark matter},
  journal = {Phys. Rev. D},
  volume = {95},
  pages = {043541},
  year = {2017},
  doi = {10.1103/PhysRevD.95.043541},
  eprint = {1610.08297},
  archivePrefix = {arXiv}
}

@article{Niemeyer2020,
  author = {Niemeyer, Jens C.},
  title = {Small-scale structure of fuzzy and axion-like dark matter},
  journal = {Prog. Part. Nucl. Phys.},
  volume = {113},
  pages = {103787},
  year = {2020},
  doi = {10.1016/j.ppnp.2020.103787},
  eprint = {1912.07064},
  archivePrefix = {arXiv}
}

@article{Ferreira2021,
  author = {Ferreira, Elisa G. M.},
  title = {Ultra-light dark matter},
  journal = {Astron. Astrophys. Rev.},
  volume = {29},
  pages = {7},
  year = {2021},
  doi = {10.1007/s00159-021-00135-6}
}

@article{PassagliaHu2022,
  author = {Passaglia, Samuel and Hu, Wayne},
  title = {Accurate effective fluid approximation for ultralight axions},
  journal = {Phys. Rev. D},
  volume = {105},
  pages = {123529},
  year = {2022},
  doi = {10.1103/PhysRevD.105.123529},
  eprint = {2201.10238},
  archivePrefix = {arXiv}
}

@article{LiuHuGrin2025,
  author = {Liu, Rayne and Hu, Wayne and Grin, Daniel},
  title = {Accurate method for ultralight axion CMB and matter power spectra},
  journal = {Phys. Rev. D},
  volume = {112},
  pages = {023513},
  year = {2025},
  doi = {10.1103/1z4c-1w7f},
  eprint = {2412.15192},
  archivePrefix = {arXiv}
}

@article{WidrowKaiser1993,
  author = {Widrow, Lawrence M. and Kaiser, Nick},
  title = {Using the Schr{\"o}dinger equation to simulate collisionless matter},
  journal = {Astrophys. J. Lett.},
  volume = {416},
  pages = {L71--L74},
  year = {1993}
}

@article{Schive2014Nature,
  author = {Schive, Hsi-Yu and Chiueh, Tzihong and Broadhurst, Tom},
  title = {Cosmic structure as the quantum interference of a coherent dark wave},
  journal = {Nature Phys.},
  volume = {10},
  pages = {496--499},
  year = {2014},
  doi = {10.1038/nphys2996},
  eprint = {1406.6586},
  archivePrefix = {arXiv}
}

@article{Schive2014PRL,
  author = {Schive, Hsi-Yu and Liao, Ming-Hsuan and Woo, Tak-Pong and Wong, Shing-Kwong and Chiueh, Tzihong and Broadhurst, Tom and Hwang, W.-Y. Pauchy},
  title = {Understanding the Core-Halo Relation of Quantum Wave Dark Matter from 3D Simulations},
  journal = {Phys. Rev. Lett.},
  volume = {113},
  pages = {261302},
  year = {2014},
  doi = {10.1103/PhysRevLett.113.261302},
  eprint = {1407.7762},
  archivePrefix = {arXiv}
}

@article{Schwabe2016,
  author = {Schwabe, Bodo and Niemeyer, Jens C. and Engels, Jan F.},
  title = {Simulations of solitonic core mergers in ultralight axion dark matter cosmologies},
  journal = {Phys. Rev. D},
  volume = {94},
  pages = {043513},
  year = {2016},
  doi = {10.1103/PhysRevD.94.043513}
}

@article{VeltmaatNiemeyer2016,
  author = {Veltmaat, Jan and Niemeyer, Jens C.},
  title = {Cosmological particle-in-cell simulations with ultralight axion dark matter},
  journal = {Phys. Rev. D},
  volume = {94},
  pages = {123523},
  year = {2016},
  doi = {10.1103/PhysRevD.94.123523},
  eprint = {1608.00802},
  archivePrefix = {arXiv}
}

@article{Mocz2017,
  author = {Mocz, Philip and Vogelsberger, Mark and Robles, Victor H. and Zavala, Jes{\'u}s and Boylan-Kolchin, Michael and Fialkov, Anastasia and Hernquist, Lars},
  title = {Galaxy formation with BECDM -- I. Turbulence and relaxation of idealized haloes},
  journal = {Mon. Not. R. Astron. Soc.},
  volume = {471},
  pages = {4559--4570},
  year = {2017},
  doi = {10.1093/mnras/stx1887}
}

@article{Veltmaat2018,
  author = {Veltmaat, Jan and Niemeyer, Jens C. and Schwabe, Bodo},
  title = {Formation and structure of ultralight bosonic dark matter halos},
  journal = {Phys. Rev. D},
  volume = {98},
  pages = {043509},
  year = {2018},
  doi = {10.1103/PhysRevD.98.043509},
  eprint = {1804.09647},
  archivePrefix = {arXiv}
}

@article{NoriBaldi2018,
  author = {Nori, Matteo and Baldi, Marco},
  title = {AX-GADGET: a new code for cosmological simulations of Fuzzy Dark Matter and Axion models},
  journal = {Mon. Not. R. Astron. Soc.},
  volume = {478},
  pages = {3935--3951},
  year = {2018},
  doi = {10.1093/mnras/sty1224},
  eprint = {1801.08144},
  archivePrefix = {arXiv}
}

@article{Mocz2019,
  author = {Mocz, Philip and Fialkov, Anastasia and Vogelsberger, Mark and Becerra, Fernando and Amin, Mustafa A. and Bose, Sownak and Boylan-Kolchin, Michael and Chavanis, Pierre-Henri and Hernquist, Lars and Lancaster, Lachlan and Marinacci, Federico and Robles, Victor H. and Zavala, Jes{\'u}s},
  title = {First Star-Forming Structures in Fuzzy Cosmic Filaments},
  journal = {Phys. Rev. Lett.},
  volume = {123},
  pages = {141301},
  year = {2019},
  doi = {10.1103/PhysRevLett.123.141301},
  eprint = {1910.01653},
  archivePrefix = {arXiv}
}

@article{MaySpringel2021,
  author = {May, Simon and Springel, Volker},
  title = {Structure formation in large-volume cosmological simulations of fuzzy dark matter: impact of the non-linear dynamics},
  journal = {Mon. Not. R. Astron. Soc.},
  volume = {506},
  pages = {2603--2618},
  year = {2021},
  doi = {10.1093/mnras/stab1764}
}

@article{Lague2024,
  author = {Lagu{\"e}, Alex and Schwabe, Bodo and Hlo{\v z}ek, Ren{\'e}e and Marsh, David J. E. and Rogers, Keir K.},
  title = {Cosmological simulations of mixed ultralight dark matter},
  journal = {Phys. Rev. D},
  volume = {109},
  pages = {043507},
  year = {2024},
  doi = {10.1103/PhysRevD.109.043507},
  eprint = {2310.20000},
  archivePrefix = {arXiv}
}

@misc{Wang2026,
  author        = {Wang, Yourong Frank},
  title         = {Lyman-$\alpha$ Forest Signatures of Mixed
                   Fuzzy and Cold Dark Matter},
  year          = {2026},
  eprint        = {2604.06038},
  archivePrefix = {arXiv},
  primaryClass  = {astro-ph.CO}
}

@article{MaBertschinger1995,
  author = {Ma, Chung-Pei and Bertschinger, Edmund},
  title = {Cosmological Perturbation Theory in the Synchronous and Conformal Newtonian Gauges},
  journal = {Astrophys. J.},
  volume = {455},
  pages = {7--25},
  year = {1995},
  doi = {10.1086/176550},
  eprint = {astro-ph/9506072},
  archivePrefix = {arXiv}
}

@article{SeljakZaldarriaga1996,
  author = {Seljak, Uro{\v s} and Zaldarriaga, Matias},
  title = {A Line-of-Sight Integration Approach to Cosmic Microwave Background Anisotropies},
  journal = {Astrophys. J.},
  volume = {469},
  pages = {437},
  year = {1996},
  eprint = {astro-ph/9603033},
  archivePrefix = {arXiv}
}

@article{LewisChallinorLasenby2000,
  author = {Lewis, Antony and Challinor, Anthony and Lasenby, Anthony},
  title = {Efficient Computation of Cosmic Microwave Background Anisotropies in Closed Friedmann-Robertson-Walker Models},
  journal = {Astrophys. J.},
  volume = {538},
  pages = {473--476},
  year = {2000},
  doi = {10.1086/309179},
  eprint = {astro-ph/9911177},
  archivePrefix = {arXiv}
}

@misc{Lesgourgues2011,
  author = {Lesgourgues, Julien},
  title = {The Cosmic Linear Anisotropy Solving System (CLASS) I: Overview},
  year = {2011},
  eprint = {1104.2932},
  archivePrefix = {arXiv},
  primaryClass = {astro-ph.IM}
}

@article{Blas2011,
  author = {Blas, Diego and Lesgourgues, Julien and Tram, Thomas},
  title = {The Cosmic Linear Anisotropy Solving System (CLASS). Part II: Approximation schemes},
  journal = {JCAP},
  volume = {07},
  pages = {034},
  year = {2011},
  doi = {10.1088/1475-7516/2011/07/034},
  eprint = {1104.2933},
  archivePrefix = {arXiv}
}

@article{LesgourguesTram2011,
  author = {Lesgourgues, Julien and Tram, Thomas},
  title = {The Cosmic Linear Anisotropy Solving System (CLASS) IV: efficient implementation of non-cold relics},
  journal = {JCAP},
  volume = {09},
  pages = {032},
  year = {2011},
  doi = {10.1088/1475-7516/2011/09/032}
}

@article{HahnAbel2011,
  author = {Hahn, Oliver and Abel, Tom},
  title = {Multi-scale initial conditions for cosmological simulations},
  journal = {Mon. Not. R. Astron. Soc.},
  volume = {415},
  pages = {2101--2121},
  year = {2011},
  doi = {10.1111/j.1365-2966.2011.18820.x},
  eprint = {1103.6031},
  archivePrefix = {arXiv}
}

@article{Almgren2013,
  author = {Almgren, Ann S. and Bell, John B. and Lijewski, Mike J. and Luki{\'c}, Zarija and Van Andel, Ethan},
  title = {Nyx: A Massively Parallel AMR Code for Computational Cosmology},
  journal = {Astrophys. J.},
  volume = {765},
  pages = {39},
  year = {2013},
  doi = {10.1088/0004-637X/765/1/39},
  eprint = {1301.4498},
  archivePrefix = {arXiv}
}

@article{AnguloHahn2022,
  author = {Angulo, Raul E. and Hahn, Oliver},
  title = {Large-scale dark matter simulations},
  journal = {Living Rev. Comput. Astrophys.},
  volume = {8},
  pages = {1},
  year = {2022},
  doi = {10.1007/s41115-021-00013-z},
  eprint = {2112.05165},
  archivePrefix = {arXiv}
}

@article{Michaux2021,
  author = {Michaux, Micha{\"e}l and Hahn, Oliver and Rampf, Cornelius and Angulo, Raul E.},
  title = {Accurate initial conditions for cosmological N-body simulations: minimizing truncation and discreteness errors},
  journal = {Mon. Not. R. Astron. Soc.},
  volume = {500},
  pages = {663--683},
  year = {2021},
  doi = {10.1093/mnras/staa3149}
}

@article{ValkenburgVillaescusa2017,
  author = {Valkenburg, Wessel and Villaescusa-Navarro, Francisco},
  title = {Accurate initial conditions in mixed dark matter--baryon simulations},
  journal = {Mon. Not. R. Astron. Soc.},
  volume = {467},
  pages = {4401--4409},
  year = {2017},
  doi = {10.1093/mnras/stx376},
  eprint = {1610.08501},
  archivePrefix = {arXiv}
}

@article{Crocce2006,
  author = {Crocce, Mart{\'i}n and Pueblas, Sebasti{\'a}n and Scoccimarro, Rom{\'a}n},
  title = {Transients from initial conditions in cosmological simulations},
  journal = {Mon. Not. R. Astron. Soc.},
  volume = {373},
  pages = {369--381},
  year = {2006},
  doi = {10.1111/j.1365-2966.2006.11040.x},
  eprint = {astro-ph/0606505},
  archivePrefix = {arXiv}
}

@article{GunnPeterson1965,
  author = {Gunn, James E. and Peterson, Bruce A.},
  title = {On the Density of Neutral Hydrogen in Intergalactic Space},
  journal = {Astrophys. J.},
  volume = {142},
  pages = {1633--1641},
  year = {1965},
  doi = {10.1086/148444}
}

@article{Rauch1998,
  author = {Rauch, Michael},
  title = {The Lyman Alpha Forest in the Spectra of Quasistellar Objects},
  journal = {Annu. Rev. Astron. Astrophys.},
  volume = {36},
  pages = {267--316},
  year = {1998},
  doi = {10.1146/annurev.astro.36.1.267}
}

@article{HuiGnedin1997,
  author = {Hui, Lam and Gnedin, Nickolay Y.},
  title = {Equation of state of the photoionized intergalactic medium},
  journal = {Mon. Not. R. Astron. Soc.},
  volume = {292},
  pages = {27--42},
  year = {1997},
  doi = {10.1093/mnras/292.1.27}
}

@article{GnedinHui1998,
  author = {Gnedin, Nickolay Y. and Hui, Lam},
  title = {Probing the Universe with the Lyman alpha forest -- I. Hydrodynamics of the low-density intergalactic medium},
  journal = {Mon. Not. R. Astron. Soc.},
  volume = {296},
  pages = {44--55},
  year = {1998},
  doi = {10.1046/j.1365-8711.1998.01249.x}
}

@article{Viel2008,
  author = {Viel, Matteo and Becker, George D. and Bolton, James S. and Haehnelt, Martin G. and Rauch, Michael and Sargent, Wallace L. W.},
  title = {How Cold Is Cold Dark Matter? Small-Scales Constraints from the Flux Power Spectrum of the High-Redshift Lyman-alpha Forest},
  journal = {Phys. Rev. Lett.},
  volume = {100},
  pages = {041304},
  year = {2008},
  doi = {10.1103/PhysRevLett.100.041304}
}

@article{Viel2013,
  author = {Viel, Matteo and Becker, George D. and Bolton, James S. and Haehnelt, Martin G.},
  title = {Warm Dark Matter as a solution to the small scale crisis: new constraints from high redshift Lyman-alpha forest data},
  journal = {Phys. Rev. D},
  volume = {88},
  pages = {043502},
  year = {2013},
  doi = {10.1103/PhysRevD.88.043502},
  eprint = {1306.2314},
  archivePrefix = {arXiv}
}

@article{PalanqueDelabrouille2015,
  author = {Palanque-Delabrouille, Nathalie and Y{\`e}che, Christophe and Baur, Julien and others},
  title = {Neutrino masses and cosmology with Lyman-alpha forest power spectrum},
  journal = {JCAP},
  volume = {11},
  pages = {011},
  year = {2015},
  doi = {10.1088/1475-7516/2015/11/011}
}

@article{Chabanier2019,
  author = {Chabanier, Sol{\`e}ne and Palanque-Delabrouille, Nathalie and Y{\`e}che, Christophe and others},
  title = {The one-dimensional power spectrum from the SDSS DR14 Lyman-alpha forests},
  journal = {JCAP},
  volume = {07},
  pages = {017},
  year = {2019},
  doi = {10.1088/1475-7516/2019/07/017},
  eprint = {1812.03554},
  archivePrefix = {arXiv}
}

@article{Arinyo2015,
  author = {Arinyo-i-Prats, Andreu and Miralda-Escud{\'e}, Jordi and Viel, Matteo and Cen, Renyue},
  title = {The Non-Linear Power Spectrum of the Lyman Alpha Forest},
  journal = {JCAP},
  volume = {12},
  pages = {017},
  year = {2015},
  doi = {10.1088/1475-7516/2015/12/017},
  eprint = {1506.04519},
  archivePrefix = {arXiv}
}

@article{Becker2011,
  author = {Becker, George D. and Bolton, James S. and Haehnelt, Martin G. and Sargent, Wallace L. W.},
  title = {Detection of extended He II reionization in the temperature evolution of the intergalactic medium},
  journal = {Mon. Not. R. Astron. Soc.},
  volume = {410},
  pages = {1096--1112},
  year = {2011},
  doi = {10.1111/j.1365-2966.2010.17507.x},
  eprint = {1008.2622},
  archivePrefix = {arXiv}
}

@article{Walther2019,
  author = {Walther, Michael and O{\~n}orbe, Jose and Hennawi, Joseph F. and Luki{\'c}, Zarija},
  title = {New Constraints on IGM Thermal Evolution from the Lyman-alpha Forest Power Spectrum},
  journal = {Astrophys. J.},
  volume = {872},
  pages = {13},
  year = {2019},
  doi = {10.3847/1538-4357/aafad1},
  eprint = {1808.04367},
  archivePrefix = {arXiv}
}

@article{Irsic2017,
  author = {Ir{\v s}i{\v c}, Vid and Viel, Matteo and Haehnelt, Martin G. and Bolton, James S. and Becker, George D.},
  title = {First Constraints on Fuzzy Dark Matter from Lyman-alpha Forest Data and Hydrodynamical Simulations},
  journal = {Phys. Rev. Lett.},
  volume = {119},
  pages = {031302},
  year = {2017},
  doi = {10.1103/PhysRevLett.119.031302},
  eprint = {1703.04683},
  archivePrefix = {arXiv}
}

@article{Armengaud2017,
  author = {Armengaud, Eric and Palanque-Delabrouille, Nathalie and Y{\`e}che, Christophe and Marsh, David J. E. and Baur, Julien},
  title = {Constraining the mass of light bosonic dark matter using SDSS Lyman-alpha forest},
  journal = {Mon. Not. R. Astron. Soc.},
  volume = {471},
  pages = {4606--4614},
  year = {2017},
  doi = {10.1093/mnras/stx1870}
}

@article{Kobayashi2017,
  author = {Kobayashi, Takeshi and Murgia, Riccardo and De Simone, Andrea and Ir{\v s}i{\v c}, Vid and Viel, Matteo},
  title = {Lyman-alpha constraints on ultralight scalar dark matter: Implications for the early and late universe},
  journal = {Phys. Rev. D},
  volume = {96},
  pages = {123514},
  year = {2017},
  doi = {10.1103/PhysRevD.96.123514}
}

@article{Nori2019,
  author = {Nori, Matteo and Murgia, Riccardo and Ir{\v s}i{\v c}, Vid and Baldi, Marco and Viel, Matteo},
  title = {Lyman-alpha forest and non-linear structure characterization in Fuzzy Dark Matter cosmologies},
  journal = {Mon. Not. R. Astron. Soc.},
  volume = {482},
  pages = {3227--3243},
  year = {2019},
  doi = {10.1093/mnras/sty2888}
}

@article{RogersPeiris2021,
  author = {Rogers, Keir K. and Peiris, Hiranya V.},
  title = {Strong Bound on Canonical Ultralight Axion Dark Matter from the Lyman-alpha Forest},
  journal = {Phys. Rev. Lett.},
  volume = {126},
  pages = {071302},
  year = {2021},
  doi = {10.1103/PhysRevLett.126.071302}
}

@article{UrenaLopez2024CLASS,
    author  = {Ure{\~n}a-L{\'o}pez, L. Arturo and Linares Cede{\~n}o, Francisco X.},
    title   = {Cosmological evolution of scalar field dark matter in the {CLASS} code: Accuracy and precision of numerical solutions},
    journal = {Physical Review D},
    volume  = {109},
    number  = {2},
    pages   = {023512},
    year    = {2024},
    doi     = {10.1103/PhysRevD.109.023512},
    eprint  = {2307.05600},
    archivePrefix = {arXiv},
    primaryClass  = {astro-ph.CO}
}

@article{Schwabe2020,
  author  = {Schwabe, Bodo and Gosenca, Mateja and
             Behrens, Christoph and Niemeyer, Jens C.
             and Easther, Richard},
  title   = {Simulating mixed fuzzy and cold dark matter},
  journal = {Physical Review D},
  volume  = {102},
  pages   = {083518},
  year    = {2020},
  doi     = {10.1103/PhysRevD.102.083518}
}

@article{MatosGuzmanUrena2000,
  author  = {Matos, Tonatiuh and Guzm{\'a}n, F. Siddhartha
             and Ure{\~n}a-L{\'o}pez, L. Arturo},
  title   = {Scalar field as dark matter in the universe},
  journal = {Classical and Quantum Gravity},
  volume  = {17},
  pages   = {1707},
  year    = {2000},
  doi     = {10.1088/0264-9381/17/7/309}
}

@article{MatosUrena2001,
  author  = {Matos, Tonatiuh and Ure{\~n}a-L{\'o}pez, L. Arturo},
  title   = {Further analysis of a cosmological model with
             quintessence and scalar dark matter},
  journal = {Physical Review D},
  volume  = {63},
  pages   = {063506},
  year    = {2001},
  doi     = {10.1103/PhysRevD.63.063506}
}

@article{Matos2009,
  author  = {Matos, Tonatiuh and V{\'a}zquez-Gonz{\'a}lez, Alberto
             and Maga{\~n}a, Juan},
  title   = {{$\phi^2$} as dark matter},
  journal = {Monthly Notices of the Royal Astronomical Society},
  volume  = {393},
  pages   = {1359--1369},
  year    = {2009},
  doi     = {10.1111/j.1365-2966.2008.13957.x},
  eprint  = {0806.0683},
  archivePrefix = {arXiv}
}

@article{Matos:1998vk,
    author = "Matos, Tonatiuh and Guzman, Francisco Siddhartha",
    title = "{Scalar fields as dark matter in spiral galaxies}",
    eprint = "gr-qc/9810028",
    archivePrefix = "arXiv",
    reportNumber = "CINVESTAV-98-22",
    doi = "10.1088/0264-9381/17/1/102",
    journal = "Class. Quant. Grav.",
    volume = "17",
    pages = "L9--L16",
    year = "2000"
}

@article{Matos:2023usa,
    author = "Matos, Tonatiuh and Ure{\~n}a-L{\'o}pez, Luis A. and Lee, Jae-Weon",
    title = "{Short review of the main achievements of the scalar field, fuzzy, ultralight, wave, BEC dark matter model}",
    eprint = "2312.00254",
    archivePrefix = "arXiv",
    primaryClass = "astro-ph.CO",
    reportNumber = "Cinvestav/2023/Fis235",
    doi = "10.3389/fspas.2024.1347518",
    journal = "Front. Astron. Space Sci.",
    volume = "11",
    pages = "1347518",
    year = "2024"
}

@article{Peng2011,
  author  = {Peng, Roger D.},
  title   = {Reproducible Research in Computational Science},
  journal = {Science},
  volume  = {334},
  pages   = {1226--1227},
  year    = {2011},
  doi     = {10.1126/science.1213847}
}

@article{Sandve2013,
  author  = {Sandve, Geir Kjetil and Nekrutenko, Anton
             and Taylor, James and Hovig, Eivind},
  title   = {Ten Simple Rules for Reproducible Computational Research},
  journal = {PLoS Computational Biology},
  volume  = {9},
  pages   = {e1003285},
  year    = {2013},
  doi     = {10.1371/journal.pcbi.1003285}
}

@article{Wilson2014,
  author  = {Wilson, Greg and Aruliah, D. A.
             and Brown, C. Titus and Chue Hong, Neil P.
             and Davis, Matt and others},
  title   = {Best Practices for Scientific Computing},
  journal = {PLoS Biology},
  volume  = {12},
  pages   = {e1001745},
  year    = {2014},
  doi     = {10.1371/journal.pbio.1001745}
}

@article{Wilkinson2016,
  author  = {Wilkinson, Mark D. and Dumontier, Michel
             and Aalbersberg, IJsbrand Jan and others},
  title   = {The FAIR Guiding Principles for Scientific
             Data Management and Stewardship},
  journal = {Scientific Data},
  volume  = {3},
  pages   = {160018},
  year    = {2016},
  doi     = {10.1038/sdata.2016.18}
}

@misc{Awais2026ULDM,
  author        = {Awais, Muhammad and Ullah, Hafiz Inam and Nunes, Rafael C. and Ferreira, Paula S. and Matos, Tonatiuh},
  title         = {Constraining Ultralight Dark Matter in Dynamical Dark Energy Cosmologies},
  year          = {2026},
  eprint        = {2610.33332},
  archivePrefix = {arXiv},
  primaryClass  = {gr-qc},
  doi           = {10.48550/arXiv.2610.33332},
  url           = {https://arxiv.org/abs/2610.33332}
}

\end{document}